\documentclass[graybox, nosecnum]{svmult}

\usepackage{mathptmx}       
\usepackage{helvet}         
\usepackage{courier}        
\usepackage{type1cm}        
\usepackage{makeidx}         
\usepackage{graphicx}        
\usepackage{multicol}        
\usepackage[bottom]{footmisc}
\usepackage{hyperref}        
\usepackage{soul}            
\hypersetup{colorlinks=true,urlcolor=blue}

\usepackage[square,numbers]{natbib}
\makeindex             
\begin{document}
\title*{X-ray Emissions from the Jovian System}
\author{W. R. Dunn}
\institute{W. R. Dunn \at Department of Physics and Astronomy, University College London, London, UK.
\at The Centre for Planetary Science at UCL/Birkbeck, Gower Street, London WC1E 6BT, UK. \email{w.dunn@ucl.ac.uk}}
%
%
\maketitle
\textit{This Chapter will appear in the Section ``Solar System Planets'' (Section Editor: Prof. G. Branduardi-Raymont) of the ``Handbook of X-ray and Gamma-ray Astrophysics'' (Editors in chief: C. Bambi and A. Santangelo)}\\
\\
\\

\abstract{The Jovian system is a treasure trove of X-ray sources: diverse and dynamic atmospheric and auroral emissions, diffuse radiation belt and Io Plasma torus emissions, and plasma-surface interactions with Io, Europa, Ganymede and Callisto. The system is a rich natural laboratory for astronomical X-rays with each region showcasing its own X-ray production processes: scattering and fluorescence of solar corona emissions; charge exchange emissions from energetic ions; Inverse-Compton, thermal and non-thermal bremsstrahlung emissions from relativistic electrons; and finger-print fluorescence lines indicative of elemental composition and the potential for life on the Galilean satellites. For the high energy astrophysics domain, perhaps Jupiter's greatest attribute is the opportunity to connect observed X-ray emissions with in-situ plasma and magnetic field measurements of the precise physical processes that lead to them - irreplaceable ground truths for systems that cannot be visited in-situ. Such simultaneous studies have revealed that Jupiter's spectacular soft X-ray flares and pulsations are produced by wave-particle interactions (compressional and electromagnetic ion cyclotron waves) and the bremsstrahlung aurorae vary with internally-driven magnetodisk reconnection and dipolarisation.  While many remote signatures remain to be linked with their source processes, the future is bright, with synergistic \textit{Chandra, NuSTAR, XMM-Newton} and \textit{Juno} in-situ measurements continuing to provide revolutionary insights in the coming years, while \textit{JUICE} and \textit{Europa} missions with \textit{ATHENA} and possibly \textit{Lynx} will enable a new legacy. However, to truly characterise some emissions (e.g. mapping Galilean satellite elemental composition) in-situ X-ray instrumentation is a necessity. Recent advances like miniaturised/Micro-Pore optics and radiation tolerant detectors enable compact, lightweight, X-ray instrumentation perfectly suited for Jupiter science. The chapter closes by reviewing feasible, low-risk concepts that would paradigm-shift our understanding of the system.}
\section{Keywords} 
X-rays, Jupiter, Aurora, Radiation Belts, Io, Europa, Ganymede, Callisto

\section{\textit{Introduction}}

\begin{figure}
    \centering
    \includegraphics[width=0.7\textwidth]{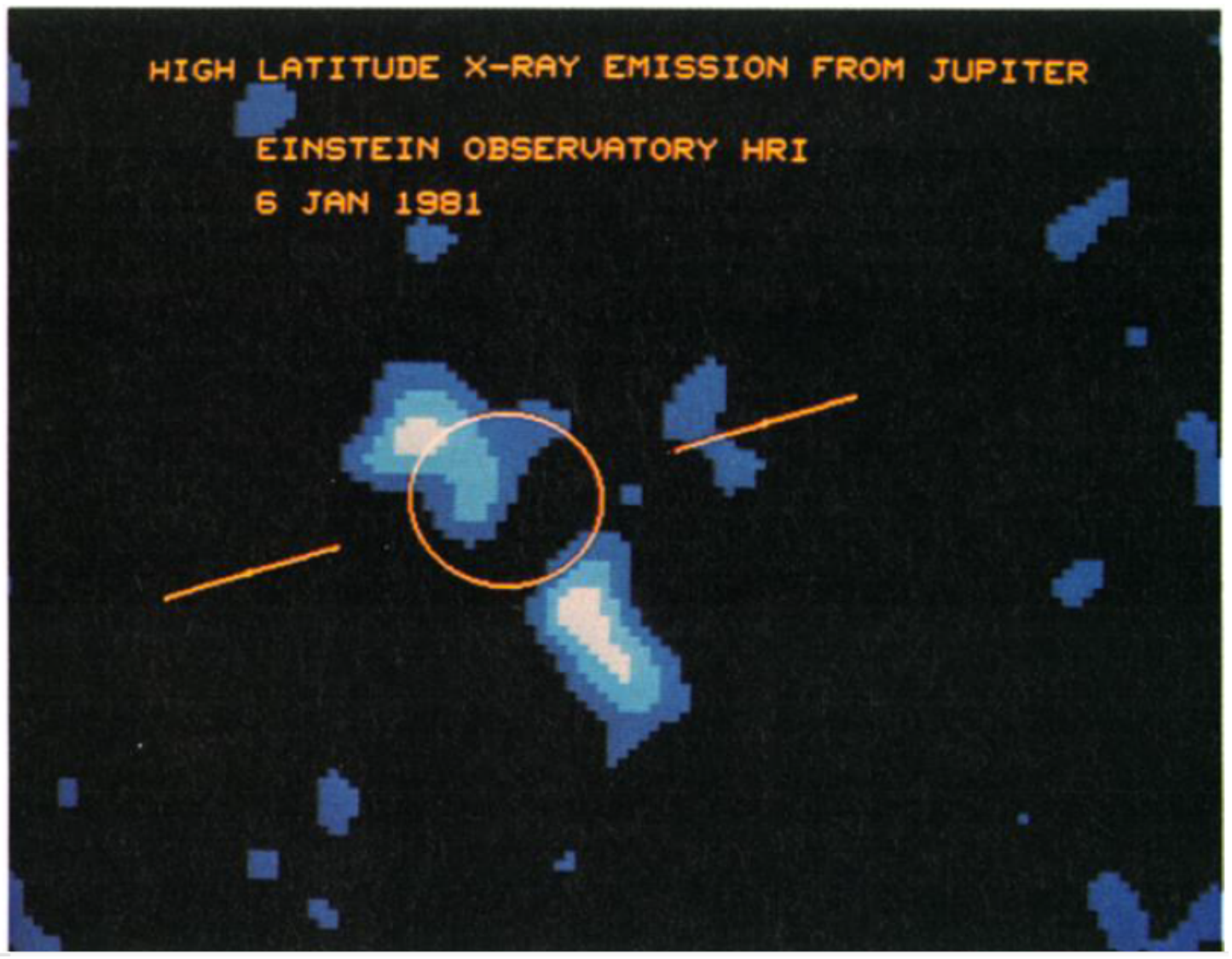}
    \caption{The first detection of X-rays from Jupiter as imaged by the Einstein Observatory in January 1981. Figure from \cite{metzger_detection_1983}.}
    \label{fig:einsteinobs}
\end{figure}

The search for X-ray emission from Jupiter began in 1962 with the \textit{Aerobee rocket} \cite{fisher_upper_1964} and continued unsuccessfully by balloon \cite{edwards_upper_1967,haymes_upper_1968,hurley_search_1972}. Expectations of X-ray bremsstrahlung emission from Earth’s aurora continued to drive searches at Jupiter, with both the \textit{Copernicus} and \textit{Uhuru} satellites conducting unsuccessful observations \cite{vesecky_upper_1975, hurley_upper_1975}. Finally, after two decades of pioneering efforts, the launch of the \textit{Einstein Observatory} ushered in the era of planetary X-ray studies, with the detection of the first X-ray emissions from another planet: Jupiter \cite{metzger_detection_1983} (Figure \ref{fig:einsteinobs}). Alongside quantifying the luminosity ($\sim10^9 W$), the \textit{Einstein Observatory's} relatively primitive spectral resolution revealed that the majority of Jupiter's emission was not, in fact, electron bremsstrahlung, as previously expected. Instead, the authors proposed that the spectral shape may be due to emission lines produced by precipitating sulphur and oxygen ions \cite{gehrels_energetic_1983} - a prediction proven a decade later.

\begin{figure}
    \centering
    \includegraphics[width=\textwidth]{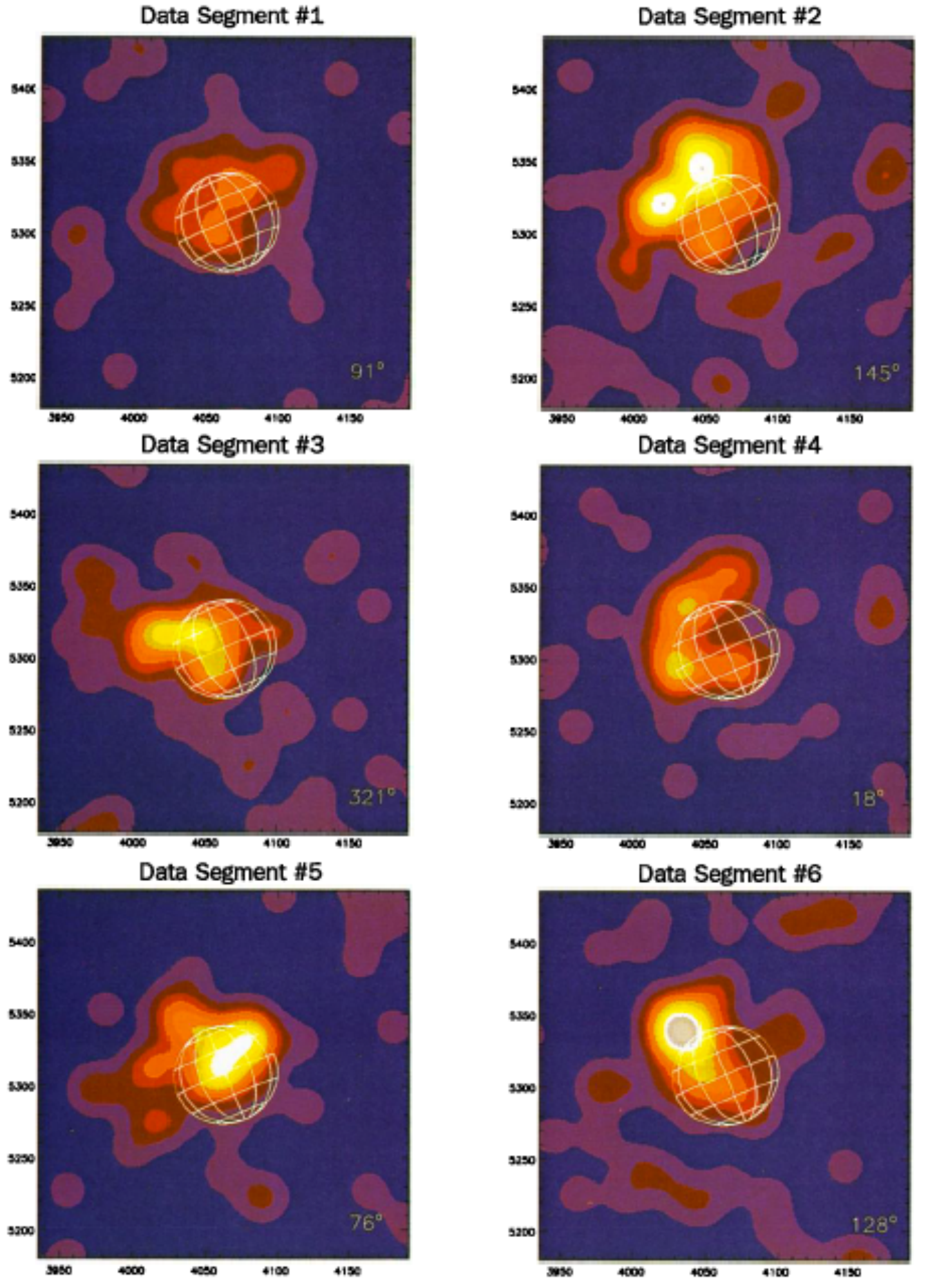}
    \caption{\textit{ROSAT High Resolution Imager} observations of Jupiter. The first evidence that Jupiter's X-ray aurora is modulated by the planet's rotation. Figure from \cite{waite_rosat_1994}.}
    \label{fig:rosataurora}
\end{figure}

During the following decade, X-ray studies of Jupiter were dormant while small-scale complementary missions became the prominent X-ray astronomy instrumentation of choice. The launch of the \textit{R{\"o}ntgen Satellite (ROSAT)} \cite{trumper_rosat-new_1993}, in April 1991, presented the ideal instrument for more detailed X-ray studies of Jupiter. \textit{ROSAT} first enabled exploration of the spatial and temporal signatures of Jupiter's X-ray emission, revealing distinct X-ray emissions from Jupiter's equatorial region and poles. Figure \ref{fig:rosataurora} shows six \textit{ROSAT} observations that first identified that the polar X-ray emissions are modulated by the planetary rotation - the aurorae rotate in and out of view with the magnetic pole of the planet, which is tilted to Jupiter's rotation axis by $\sim10^\circ$. \textit{ROSAT} also revealed that the collision of the comet Shoemaker-Levy 9 with Jupiter appeared to trigger bright X-ray emissions from the planet \cite{waite_rosat_1995}.

\begin{figure}
    \centering
    \includegraphics[width=0.9\textwidth]{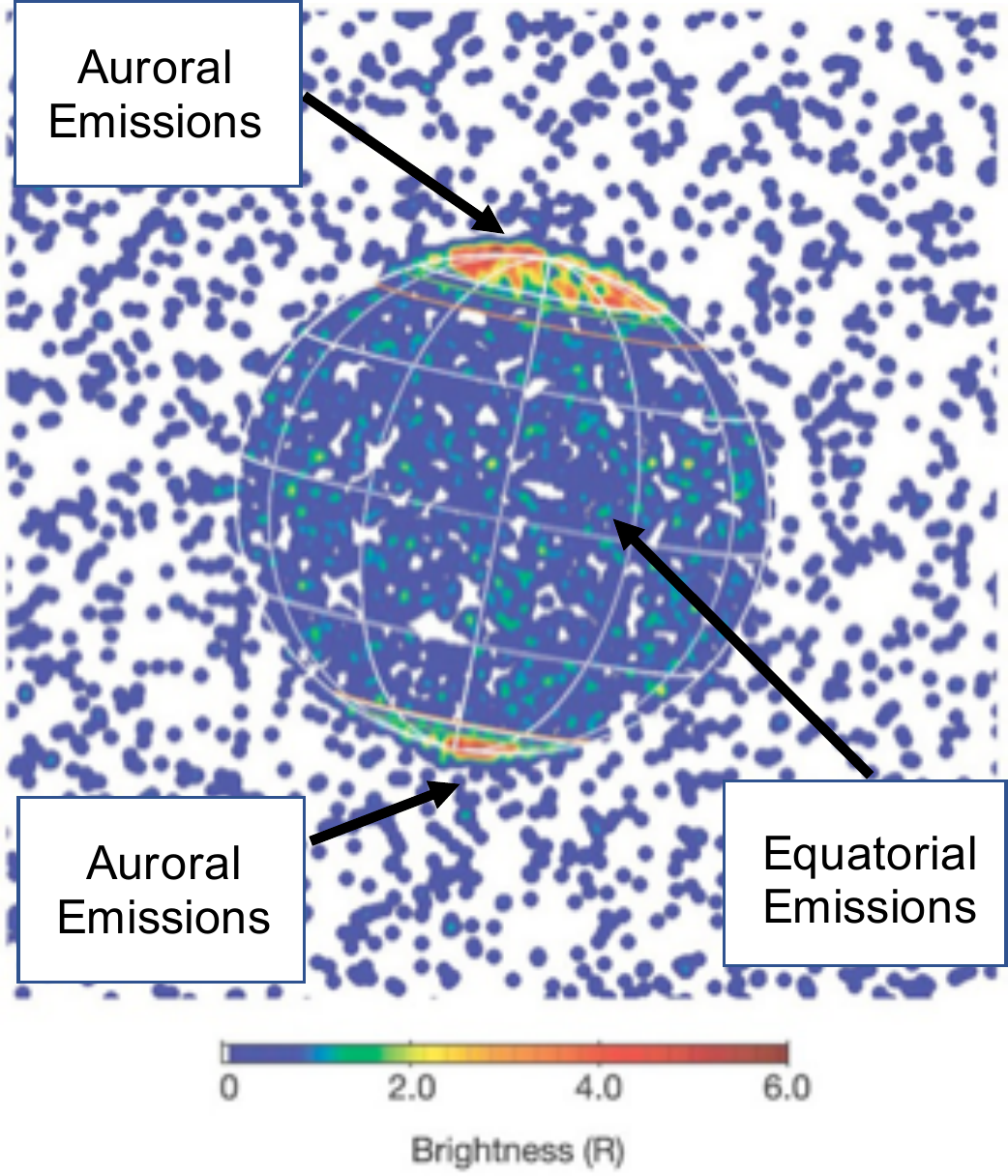}
    \caption{Taken in December 2000, this is the first Chandra HRC X-ray image of Jupiter. The colourbar indicates brightness in rayleighs (R). X-ray photons have been smeared by double the 0.4'' FWHM PSF of \textit{Chandra's HRC} instrument.  Jupiter is overlaid with a   30° latitude-longitude graticule (white).
    \cite{gladstone_pulsating_2002}. Labels have been added to indicate Jupiter's auroral and equatorial emissions. Original figure from \cite{gladstone_pulsating_2002}.}
    \label{fig:gladstone2002Jupiter}
\end{figure}

\begin{figure}
    \centering
    \includegraphics[angle =270, width=\textwidth]{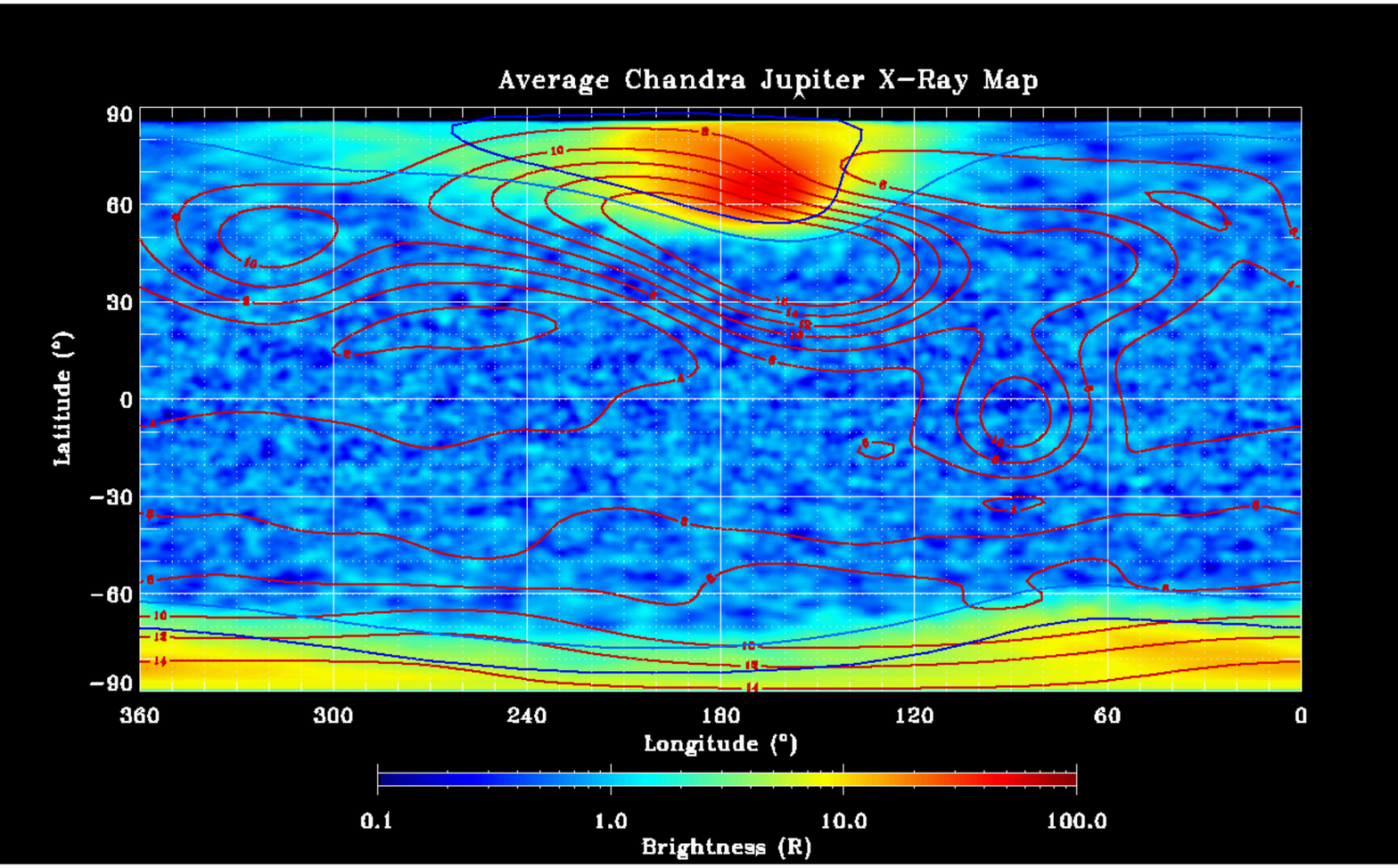}
    \caption{System III Longitude-Latitude X-ray Maps of Jupiter combining \textit{Chandra} observations of the planet between 2017-2019. The outer blue contour indicates the footprint of Jupiter’s satellite Io as it orbits the planet. Magnetic field lines in this region map to an equatorial location that is 5.9 R$_J$ (Jupiter radii) from the planet. The inner blue contour shows the location of Jupiter’s main auroral oval (mapping to an equatorial location of 20-50 R$_J$). The red contours show the surface magnetic field strength at Jupiter’s surface as measured by \textit{Juno}. The numerical values on these red contours show the magnetic field strength in Gauss \cite[Credit: G. R. Gladstone]{CXCNews27}.}
    \label{fig:gladstonemap}
\end{figure}

In 1999, the launch of \textit{XMM-Newton}\cite{jansen_xmm-newton_2001} and \textit{Chandra}\cite{weisskopf_chandra_2000} ignited a new era in X-ray astronomy, and the two observatories continue to revolutionise our understanding of Jupiter today. These highly complementary instruments introduced significant improvements over their predecessors. Alongside both observatories' high time resolution ($<3$s), \textit{Chandra's} unrivalled X-ray spatial resolution (0.5") enables highly resolved X-ray maps of Jupiter - tracking the variation in X-rays at a given planetary longitude and latitude with time. Figure \ref{fig:gladstone2002Jupiter} and \ref{fig:gladstonemap} highlight \textit{Chandra High Resolution Camera (HRC)} Jupiter observation capabilities, showing the first \textit{Chandra} X-ray observation of Jupiter and a longitude-latitude map of averaged X-ray brightness across Jupiter.

\begin{figure}
    \centering
    \includegraphics[width=0.9\textwidth]{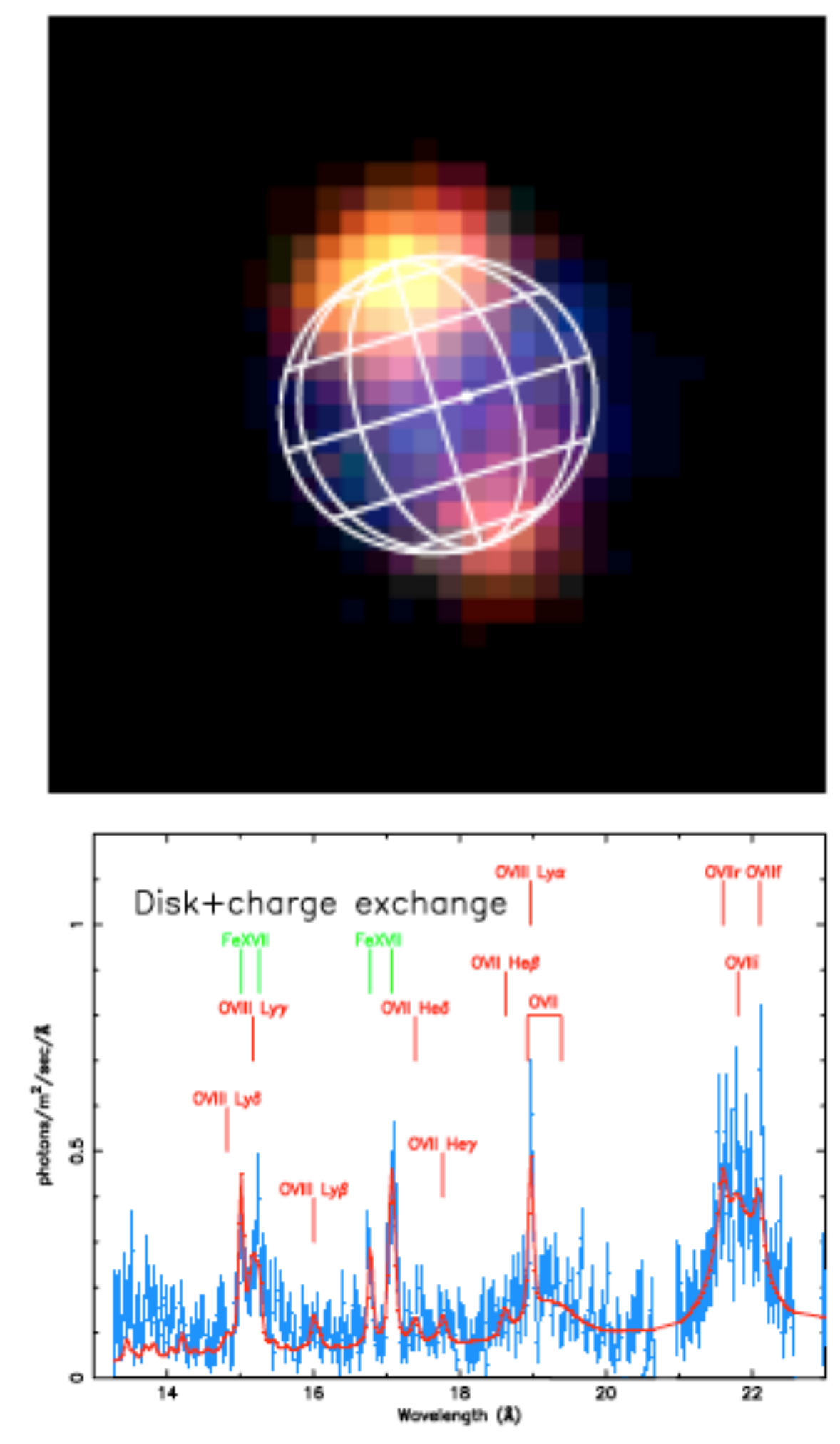}
    \caption{Top panel: 2.9-pixel-smoothed EPIC-pn image of Jupiter; North is to the top and East to the left. Colours show different energy ranges with 0.2-0.5 keV (red), 0.5-0.7 keV (green) and 0.7-2.0 keV (blue). The dot shows the sub-solar point, with the equatorial emission slightly shifted towards this. The sub-Earth point is at the centre of the graticule. 
    Lower panel:
    Combined RGS1 and 2 Jupiter spectra (blue crosses). Overlaid in red is the best fit model, which includes both a charge exchange model, representative of Jupiter's soft X-ray aurorae, and solar X-ray lines to represent the Jovian disk emission from scattered photons from the solar corona. Top panel from \cite{branduardi-raymont_first_2004}. Lower panel from \cite{branduardi-raymont_study_2007}.}
    \label{fig:gbr2004xmm}
\end{figure}

\textit{XMM-Newton} provides exceptional spectral resolution (E/$\Delta$E $\sim$ 100-500 for the RGS instrument and 20-50 for EPIC) and heightened sensitivity. For Jupiter, \textit{XMM-Newton} collects $\sim$3-8 times more photons per unit time than \textit{Chandra} - enabling detailed spectral studies to characterise the emission and sufficiently high-signal short-timescale (few-minute - time resolution limited by signal) temporal signatures for comparison with in-situ time series data. The top panel of Figure \ref{fig:gbr2004xmm} shows a spatially and spectrally resolved X-ray image of Jupiter with \textit{XMM-Newton's EPIC-pn} instrument, coloured by energy. The lower panel of the figure showcases the high resolution spectra available using \textit{XMM-Newton's Reflection Grating Spectrometer (RGS)}.

For planetary studies, a key attribute of the \textit{Chandra} and \textit{XMM-Newton} observatories is their long ($\sim$ 40 hour) orbital period which enables planetary observations for 10s of hours continuously. The opportunity to remotely track variation at minute timescales for multiple planetary rotations (a Jupiter day is 9.925 hours) has proven invaluable for Jupiter studies, isolating cause and effect by ensuring variation is tracked over suitable timescales and can be connected with the in-situ processes responsible.

Alongside studies of Jupiter itself, X-ray observatories have enabled the remote study of many other structures and bodies in the Jupiter system. \textit{Chandra's} capabilities have opened the door to X-ray studies of Jupiter's four largest moons, the Galilean satellites: Io, Europa, Ganymede and Callisto \cite{elsner_discovery_2002,nulsen_x-ray_2020}. The very first \textit{Chandra} observations of Jupiter also directly imaged the ring of plasma encircling Jupiter that is produced by Io's volcanoes - the Io Plasma Torus \cite{elsner_discovery_2002}. Furthering the study of the space plasma surrounding Jupiter, the \textit{Suzaku} satellite \cite{mitsuda_x-ray_2007} has laid essential foundations for remote imaging of Jupiter's radiation belts, repeatedly observing this structure \cite{ezoe_discovery_2010,numazawa_suzaku_2019,numazawa_suzaku_2021}.

In 2012, X-ray observing capabilities were further expanded with the launch of NASA's \textit{NuSTAR} mission\cite{harrison_nuclear_2013}, which has provided revolutionary insights into the highest energy X-ray emissions observed from planetary bodies, characterising Jupiter's X-ray emissions up to 20 keV \cite{Mori_NuSTAR}. This has further revealed the extreme energetics of Jupiter's aurorae. 

For the high energy astrophysics domain generally, arguably the greatest advantage of solar system bodies is that they can be visited by in-situ spacecraft. This presents the novel opportunity to connect remotely observed X-ray signatures with measurements of the precise physical processes that generate them; irreplaceable ground-truths for systems beyond our in-situ reach. Since the very first X-ray observations of Jupiter, such synergies between in-situ spacecraft and remote X-ray observations have been utilised. The 1979 \textit{Einstein} X-ray observations occurred contemporaneously with the first \textit{Voyager} flyby of Jupiter \cite{krimigis_characteristics_1981}. The \textit{Galileo} spacecraft Jupiter orbits from December 1995 to September 2003 \cite{bagenal_jupiter_2007} were coincident with both \textit{ROSAT} observations \cite{gladstone_secular_1998} and the first \textit{Chandra} \cite{gladstone_pulsating_2002,elsner_discovery_2002,elsner_simultaneous_2005} and \textit{XMM-Newton} observations \cite{branduardi-raymont_first_2004,branduardi-raymont_study_2007,branduardi-raymont_latest_2007}. The December 2000 \textit{Chandra} observations also represent the only remote observations supported by multiple in-situ spacecraft, with the \textit{Cassini spacecraft's} approach towards the planet, providing solar wind measurements upstream of the planet simultaneous with in-situ magnetosphere measurements by \textit{Galileo}\cite{hospodarsky_simultaneous_2004}. In February and March 2007, a challenging campaign of relatively short \textit{Chandra} and \textit{XMM-Newton} observations were taken coincident with \textit{New Horizons} approach towards Jupiter, enabling correlations between auroral emissions and the solar wind conditions \cite{dunn_comparisons_2020}. Most recently, and ongoing at the point of writing, we are fortunate to have the \textit{Juno} spacecraft\cite{bolton_jupiters_2017} taking in-situ measurements at Jupiter simultaneous with \textit{Chandra}, \textit{XMM-Newton} and \textit{NuSTAR} campaigns \cite[for example]{weigt_chandra_2020, wibisono_jupiters_2021,wibisono_temporal_2020,yao_revealing_2021}. \textit{Juno} is equipped with a suite of plasma \cite{mauk_juno_2017}, magnetic field \cite{connerney_juno_2017} and remote instrumentation \cite{kurth_juno_2017,gladstone_ultraviolet_2017} that is ideal for studying the processes that lead Jupiter's magnetosphere to generate X-ray emissions. While \textit{Juno} does not carry an X-ray instrument on board, the synergies enabled by simultaneous X-ray observatory and Juno measurements are proving ground-breaking and will continue to be transformative for years to come.

Alongside in-situ measurements, comparison with other wavebands, such as the UV coverage provided by the \textit{Hubble Space Telescope (HST)}, has also proven to be key for constraining the physical processes associated with X-ray emissions. Simultaneous \textit{Chandra} and \textit{HST} observations in February 2003 played a key role in characterising UV-X-ray auroral connections \cite{branduardi-raymont_spectral_2008,elsner_simultaneous_2005}. More recently, an extensive campaign of HST observations scheduled to coincide with the \textit{Juno} mission \cite{grodent_jupiters_2018} have provided a rich new catalogue of $\sim$30 HST observations that are simultaneous with either \textit{Chandra} or \textit{XMM-Newton} observations. These provide an extensive dataset that has only just begun to be explored \cite{wibisono_jupiters_2021}.

For practicalities of analysis, a wide-range of bespoke tools have been used to study Jupiter's X-ray emissions. From the  open-source toolkits built for X-ray astronomy, the \textit{XMM-Newton} and \textit{Chandra} data is often processed with the \textit{SAS} and \textit{CIAO} tools built for each observatory respectively. Over the past two decades, Jovian X-ray spectral model fitting has most often been performed with the XSPEC \cite{arnaud_xspec_1996} package, which is built to handle \textit{XMM-Newton} and \textit{Chandra} X-ray spectra.

This chapter is organised by structures within the Jovian system. We present the current (as of February 2022) understanding of X-ray emissions from Jupiter's: equatorial regions, aurorae, Io Plasma Torus, radiation belts and from the Galilean satellites. Finally, we close by looking forward to future opportunities and instrumentation.

\section{\textit{Jupiter's Equatorial Emissions}}    \label{section:eqr}

\textit{ROSAT} observations first differentiated Jupiter's equatorial and auroral emissions \cite{waite_equatorial_1997}, and revealed that the equatorial emission occurred predominately in locations on which sunlight was incident\cite{gladstone_secular_1998}. Following this finding,  models showed that solar-photon driving was sufficient to reproduce the majority of the observed equatorial X-ray brightness \cite{maurellis_jovian_2000}. Because H$_2$ is the most abundant molecule in the Jovian atmosphere, Thomson (elastic) scattering of solar photons is the dominant process, while k-shell fluorescence (e.g from carbon in hydrocarbons) contributes less to the total emission (8-12$\%$). Saturn's H$_2$-dominated atmosphere demonstrates a similar trend \cite{bhardwaj_chandra_2005}, while e.g. Venus is fluorescence-dominated due to the relative abundances of complex molecules in its atmosphere \cite{cravens_x-ray_2001}.

\begin{figure}
    \centering
    \includegraphics[width=0.8\textwidth]{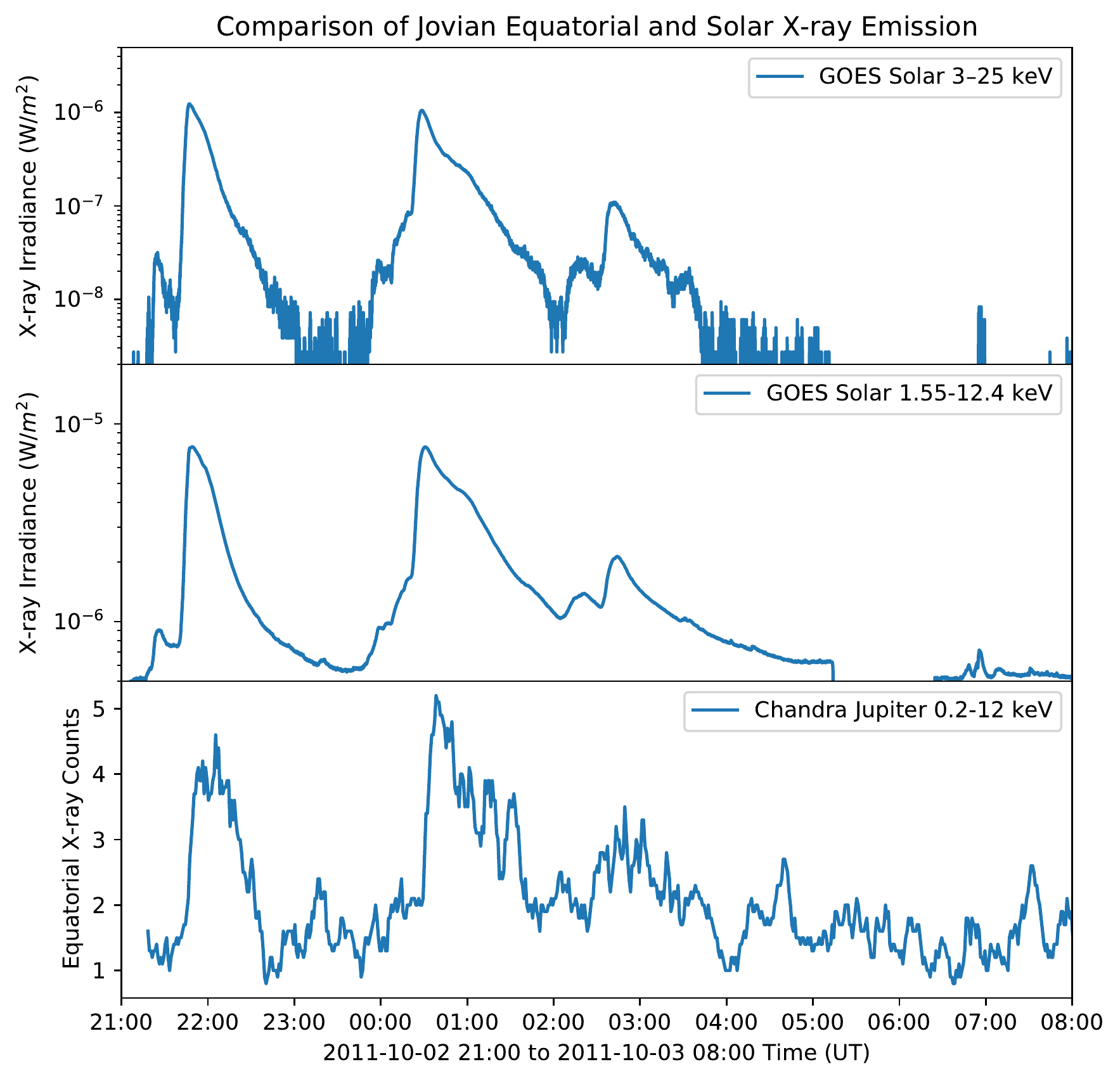}
    \caption{Three panels comparing \textit{GOES} observations of the Solar X-ray Irradiance in the 3-25 keV (top panel) and 1.55-12.4 keV (middle panel) bands and the 5-minute moving mean X-ray photon counts from Jupiter's equatorial region (bottom panel), as observed by the \textit{Chandra ACIS} instrument from 2nd to 3rd of October 2011. The Jupiter observations have been shifted to account for the light travel time from \textit{GOES} to the Sun, then the Sun to Jupiter and Jupiter to Chandra.}
    \label{fig:GOESandJupiterEqr}
\end{figure}

Since the \textit{ROSAT} era, observations by \textit{Chandra} and \textit{XMM-Newton} have shown that there is a strong link between Jupiter's equatorial emissions and the solar X-ray flux \cite{branduardi-raymont_latest_2007, bhardwaj_solar_2005}. Jupiter's entire sunlit disk varies in brightness with solar irradiance on timescales of single observations, but this is most prominent in the equatorial regions. Figure \ref{fig:GOESandJupiterEqr} shows how the equatorial emission reproduces the impulsive brightening and duration of solar flares observed by the GOES spacecraft. Over the Sun's $\sim$ 11-year activity cycle, Jupiter's equatorial X-ray emission can go from a bright well-defined disk at solar maximum (such as that in Figure \ref{fig:gladstone2002Jupiter}) to being difficult to distinguish from the sky background at solar minimum \cite{dunn_jupiters_2020}. 

\begin{figure}
    \centering
    \includegraphics[width=0.9\textwidth]{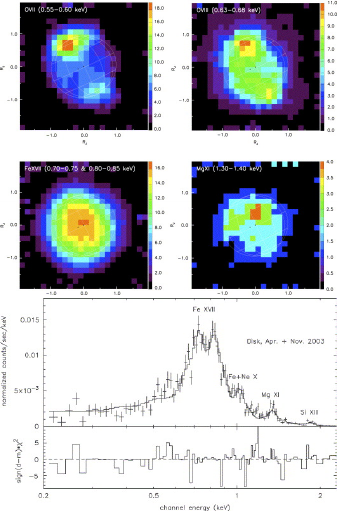}
    \caption{Upper 4 panels: Smoothed XMM-Newton EPIC images of Jupiter in energy ranges dominated by specific spectral lines: 0.55–0.60 keV O VII emissions (upper left), 0.63–0.68 keV O VIII emissions (upper right), 0.70–0.75 and 0.80–0.85 Fe XVII emissions (lower left) and 1.30–1.40 keV Mg XI emissions (lower right). X-ray counts are shown on the colour bar. A $30^\circ$ latitude-longitude graticule is overlaid on Jupiter in grey. The sub-solar point is shown by a circular dot, while the graticules are centered on the sub-Earth point. Lower 2 panels:  the upper panel shows combined \textit{XMM-Newton EPIC-pn} observations from April and November 2003 of Jupiter's equatorial disk spectrum (crosses) overlaid with the best fit model (line). Labels on the spectrum indicate different spectral lines from Fe XVII, a blend of FeXXI and Ne X, and lines from Mg XI and Si XIII.
    The lower panel shows the model as the horizontal dashed line at zero, with contributions of the spectrum away from the model shown as the histograms above and below the line. Figures from \cite{branduardi-raymont_latest_2007}.}
    \label{fig:GBR2007Disk}
\end{figure}

\begin{figure}
    \centering
    \includegraphics[width=0.75\textwidth]{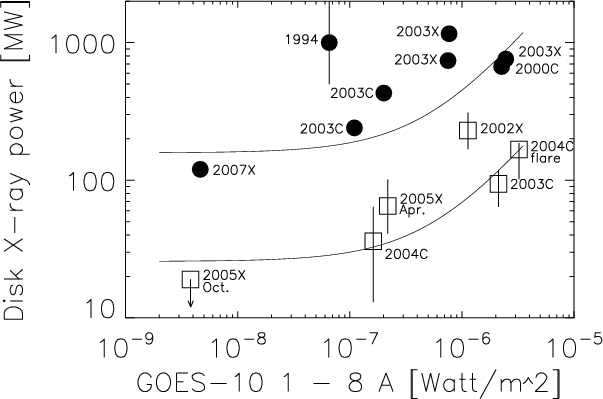}
    \caption{Scatter plot showing the X-ray power from Jupiter's (circles) and Saturn's (squares) disks (0.2-2.0 keV) plotted against the solar (1.5-12.4 keV) X-ray irradiance measured by the \textit{GOES-10 satellite}, for \textit{Chandra} (C), \textit{XMM-Newton} (X) and \textit{ROSAT} (1994) observations. The continuous lines are linear square fits. Figure from \cite{branduardi-raymont_x-rays_2010}.}
    \label{fig:GBR2010Disk}
\end{figure}

Spectrally, at solar maximum, Jupiter's equatorial emissions are dominated by Fe-lines that peak at 0.7-0.9 keV, with further contributions from Ne and Mg up to 1.5 keV (shown in Figure \ref{fig:GBR2007Disk}). During the solar declining phase and solar minimum, the peak of the spectrum shifts to lower energies and the spectrum is reduced in height by lower photon fluxes \cite{dunn_jupiters_2020}. Such low count rates make it more challenging to model at minimum, but solar corona spectral models have been found to offer a good representation of Jupiter's equatorial emission across the solar cycle.  Although, sometimes best-fit models do require the addition of more enhanced spectral lines of e.g. Mg at 1.3-1.4 keV as well as the coronal model \cite{bhardwaj_low-middle-latitude_2006,branduardi-raymont_latest_2007}. Modelling of scattered solar emissions with a Jovian albedo of $10^{-4}-10^{-2}$ were able to reproduce observations up to 0.8 keV, but found discrepancies for energies higher than this, implying required updates to coronal models \cite{cravens_x-ray_2006}.

Typical examples of models used to reproduce the coronal emission include the MEKAL \cite{mewe_update_1995} or Astrophysical Plasma Emission Code (APEC - \cite{smith_collisional_2001}), with solar abundances and the correct temperature and photon fluxes. The required temperatures for the plasma vary with solar cycle, with kT temperatures of 0.4-0.65 keV at solar maximum, 0.2-0.3 keV in the solar declining phase and 0.1-0.2 keV at solar minimum \cite{bhardwaj_low-middle-latitude_2006,branduardi-raymont_latest_2007, dunn_jupiters_2020}. For solar maximum (minimum), the resulting equatorial X-ray spectrum peaks at higher (lower) energies and has higher (lower) photon fluxes. Consequently, Figure \ref{fig:GBR2010Disk} shows that the measured power of Jupiter's equatorial emissions can vary from $\sim$0.1 GW at solar minimum to $\sim$1 GW at solar maximum\cite{branduardi-raymont_x-rays_2010, dunn_jupiters_2020}. 

Interestingly, the spectral peak of solar emission observed through the scatter from Jupiter's upper atmosphere is below the energy limit of the GOES observatories (e.g. Figure \ref{fig:GOESandJupiterEqr}). While this presents the opportunity for new insights into solar physics through studies of the Jovian emission, one should also be cautious when propagating the GOES solar X-ray fluxes to the gas giants: we can expect substantial, and currently uncharacterised, uncertainties between propagated GOES fluxes and the solar-driven emissions of the gas giants.

It should be noted that while we have labelled the scattered solar emission as `equatorial', in order to distinguish it from the auroral emissions, it is in fact observed across the sunlit disk of Jupiter and will also contribute some low-level emission in the auroral zone. 

\subsubsection{Possible Additional Equatorial Contributions}

While there is a substantial body of evidence to suggest that the equatorial emissions are dominated by scattered and fluoresced solar photons \cite{bhardwaj_chandra_2005, bhardwaj_solar_2005, maurellis_jovian_2000, branduardi-raymont_latest_2007, dunn_jupiters_2020}, this may not be the only process producing X-ray emissions from the planet's atmosphere. The limited resolution of the \textit{ROSAT} observations initially sparked a different proposed origin for the emission, and while this is no-longer expected to be the dominant source of emission, the possibility of additional equatorial emission processes has persisted. 

\begin{figure}
    \centering
    \includegraphics[width=0.75\textwidth]{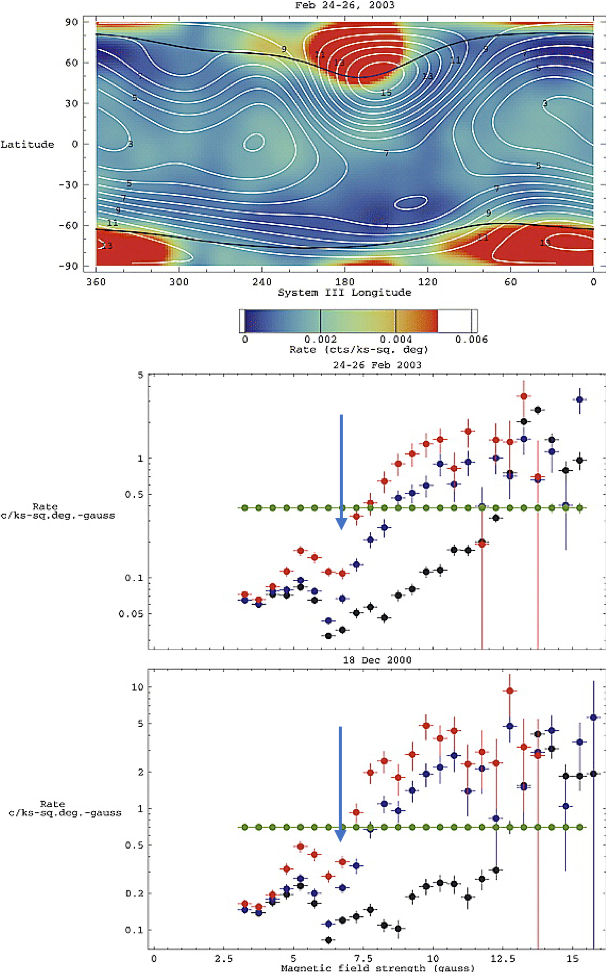}
    \caption{Upper Panel: X-ray rate-map combining \textit{Chandra ACIS-S} and \textit{HRC-I} exposures from 24-26 February 2003 Chandra data, in Jupiter's SIII coordinates. X-ray emissions have been convolved with a two-dimensional Gaussian with $\sigma = 10^\circ$. The white contours show the surface magnetic field strength in Gauss and the black lines show the magnetic footprints of Io, as defined by the VIP4 model\cite{connerney_new_1998}. Counts per kilosecond per square degree are shown with the colour bar, with an intensity scale limited to 0.15 times the maximum auroral rate in order to highlight low latitude structures. To limit the impact of uncertainties in the projected location of X-rays, only events more than $30^\circ$ longitude from the limb were included. Middle and Lower Panel: distribution of count rates vs surface magnetic field strength for (middle) the 24-26 February 2003 \textit{Chandra ACIS-S} and \textit{HRC-I} observations and (lower) the 18 December 2000 \textit{Chandra HRC} observation. Shown on the plots are data for the full disk (black), latitudes between -$45^\circ$ to $45^\circ$ (blue), and -$30^\circ$ to $30^\circ$ (red), uniformly distributed data with the same number of total counts (green). Blue arrows highlight the dip in emission at higher surface field strengths in the equatorial zone. Figures from \cite{bhardwaj_low-middle-latitude_2006}.}
    \label{fig:Bhardwaj2006}
\end{figure}

Observational hints by \textit{ROSAT} that the equatorial emissions are organised by surface magnetic field strength prompted the possibility that there may be direct precipitation of particles from Jupiter's intense radiation belts into the atmosphere\cite{waite_equatorial_1997,gladstone_secular_1998}. Follow-up studies by Chandra provided further hints that Jupiter's X-ray emission may be organised by surface magnetic field strength \cite{bhardwaj_low-_2006}. Figure \ref{fig:Bhardwaj2006} shows a count-rate map for Jupiter, with increases in equatorial emission appearing to coincide with the regions of low surface field strength. The lower two panels of the figure show X-ray count-rate (c/ks-sq.deg-gauss) versus surface magnetic field strength (gauss). At higher surface field strengths (coincident with the poles), the aurorae dominate the count-rates. However, for multiple observations and instruments, the non-auroral zones appear to show heightened emission for surface field strengths below 6 G, with a dip at slightly higher field strengths (see blue arrow). Low count-rates unfortunately prevented the authors from comparing the spectrum for different surface magnetic field strengths - and thus investigating potentially different source processes - but they note that there appears to be an excess of 1.3 keV (potentially Mg) emission in this region \cite{bhardwaj_low-_2006}.

\begin{figure}
    \centering
    \includegraphics[width=\textwidth]{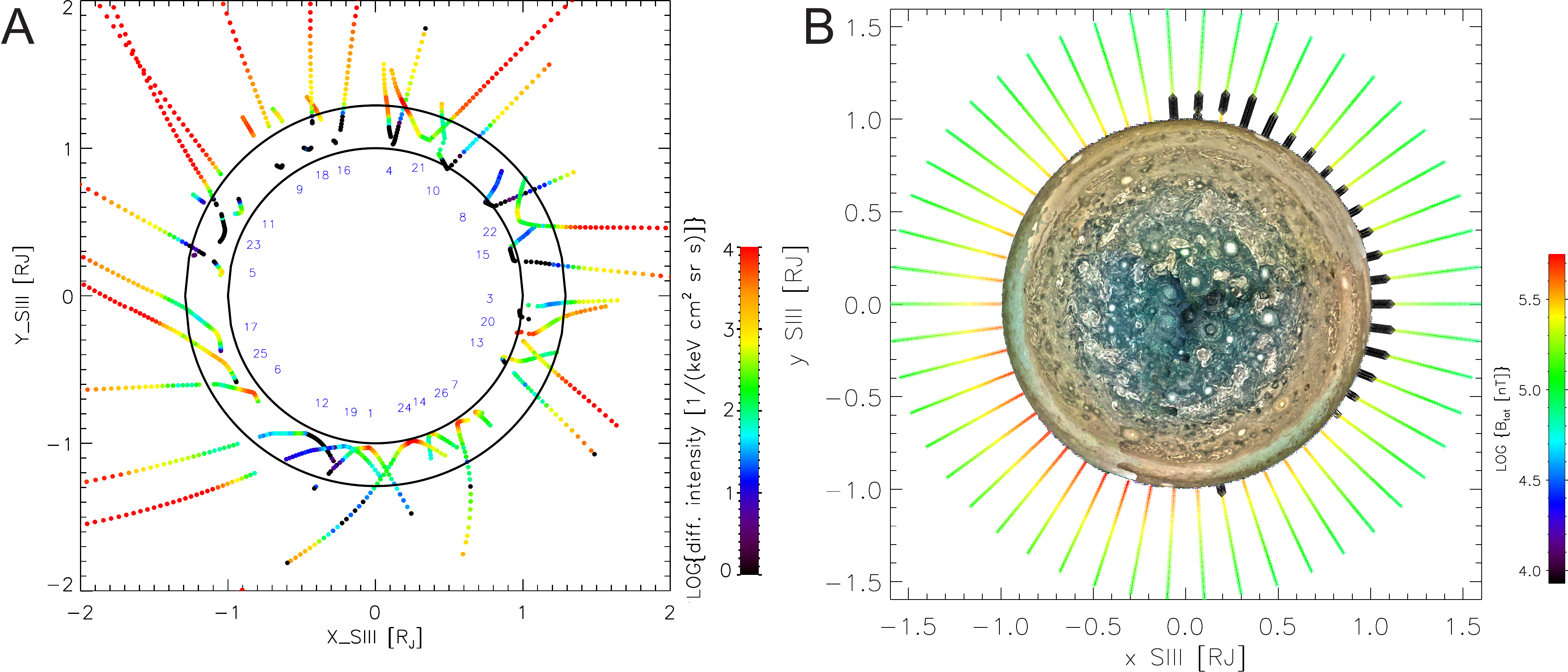}
    \caption{Left panel: Measurements of the 1 MeV total particle intensity (color coded) from the \textit{Juno JEDI} instrument, filtered for measurements $\>10^\circ$ from the loss cone. The measurements are projected onto the xy-plane of the right-handed (east) SIII coordinate system. Jupiter's surface is shown by the inner black circle, while the main ring is shown by the outer black circle. Juno orbit numbers are shown  with blue numbers. The figure shows that the innermost radiation belt (red areas near inner circle) was not always detected. Right panel: Magnetic field strength at the magnetic equator, shown at the location where the respective field line cuts through the rotational equatorial plane. Black diamonds indicate regions in the rotational equatorial plane where no stable magnetic trapping is possible. Figure from \cite{kollmann_jupiters_2021}, credit: NASA/SwRI/MSSS/Gervasio Robles.}
    \label{fig:kollmannradbelt2021}
\end{figure}

The proposed explanation for these localised enhancements is that regions of low surface magnetic field strength have larger loss cones and permit otherwise trapped energetic radiation belt ions and electrons to precipitate directly into the upper atmosphere, where collisions with the neutral population would produce X-ray emissions (associated charge exchange and/or bremsstrahlung processes are detailed in later sections). Indeed, measurements by the Juno spacecraft do suggest that certain magnetic field regions close to the planet may not be able to trap particles (Figure \ref{fig:kollmannradbelt2021}) \cite{kollmann_jupiters_2021}. However, the map in Figure \ref{fig:gladstonemap} does not show any clear preference for concentrations of emission in low surface magnetic field location. A careful study that excludes intervals of high solar X-ray flux, which would otherwise dominate the X-ray emission, is likely to be necessary to test for the presence of direct radiation belt precipitation, and/or other X-ray-production processes.

\section{\textit{Jupiter's X-ray Aurorae}} 
    \label{sec:aurora}

Aurorae are spectacular displays of light produced at a body's magnetic poles. They are created by the interaction between the magnetosphere of the body (the cavity around a body that is governed by its magnetic field) and the atmosphere. These dancing light displays provide invaluable diagnostics of the fundamental plasma processes governing the space environment surrounding a body. Jupiter has the most powerful aurorae in the solar system.  

In order to interpret the X-ray emissions from Jupiter's aurorae, it is important to introduce context for these emissions through a short orientation of Jupiter's magnetosphere. For brevity, we only provide a brief overview of those aspects relevant to this chapter. We encourage interested readers to explore more detailed reviews \cite{bagenal_jupiter_2007, badman_auroral_2015, bagenal_magnetospheric_2014}.

Many have said that Jupiter is `the planet of superlatives'. The Jovian magnetosphere is no exception to this rule. Jupiter's magnetosphere is the largest coherent structure in the heliosphere.  Sunwards, the magnetopause (the boundary between the magnetosphere and the solar wind) has a standoff distance that varies bi-modally between 63 and 92 $R_J$ (1 $R_J$ = 1 Jupiter Radius) \cite{joy_probabilistic_2002}. This bimodal state is produced by compressions (expansions) of the magnetosphere from high (low) solar wind ram pressure. Anti-Sunwards, the tail of the magnetosphere stretches 5 AU from Jupiter, reaching the orbit of Saturn\cite{weigt_searching_2021}. This vast magnetic cavity is produced through the complex interactions between Jupiter's rapid rotation (a Jupiter day is 9.925 hours), its strong magnetic field (surface field strengths of $\sim$ Gauss - Figure \ref{fig:gladstonemap}) and the 1 ton/s injections of material (predominantly $SO_2$) from Io's volcanoes, of which $250$ kg s$^{-1}$ becomes dissociated and ionised to contribute plasma to the system \cite{bagenal_magnetospheric_2014}.

\begin{figure}
    \centering
    \includegraphics[width=0.75\textwidth]{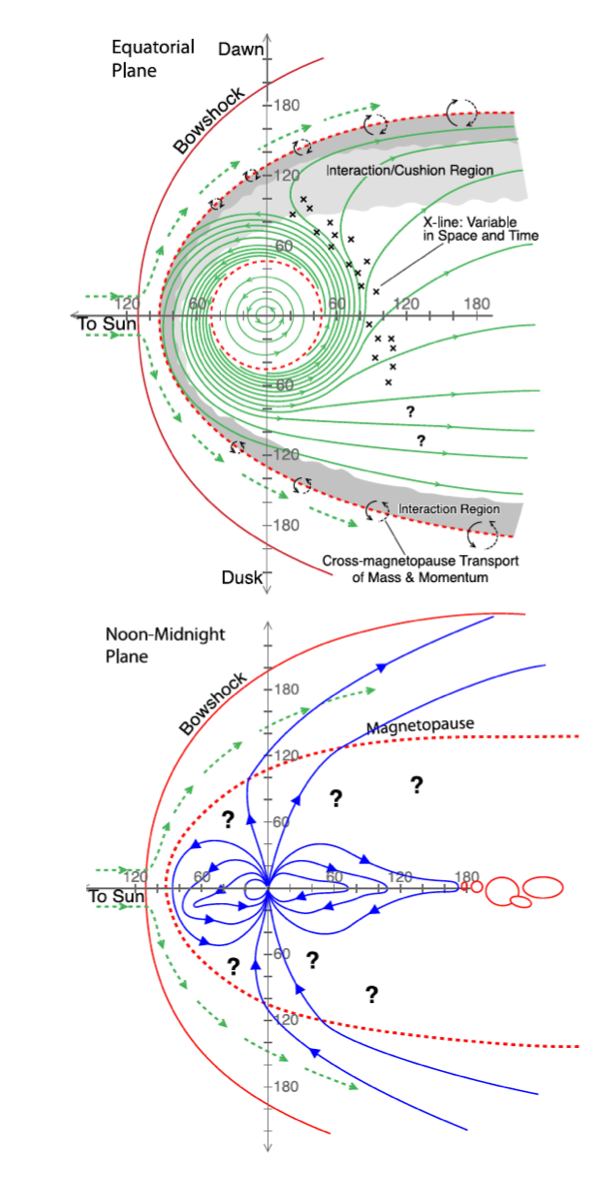}
    \caption{The top panel shows a top-down view of Jupiter's magnetospheric structures and dynamics in the equatorial plane, while the lower panel shows it as observed in the noon-midnight plane. Green lines indicate the flow of plasma. Red lines show boundaries between different plasma regimes. Blue lines show magnetic field lines. Inside 50 $R_J$, plasma is close to corotation with the planet. Beyond this, the plasma significantly lags corotation. Field lines become bent and the plasma spirals out as the radial flow becomes comparable to the azimuthal flow. Centrifugal forces lead the plasma to form a disk. This dense equatorial plasma applies tension to the field lines in the night side and stretches them (see blue field lines in lower panel). When these field lines become sufficiently stretched, the current sheet that separates the oppositely directed lines is no longer sufficiently thick to stop them coming into contact and magnetic reconnection occurs (along the lines labelled with x symbols in upper panel). This results in plasmoid release (red circles - lower panel). Once emptied, to conserve flux the post-reconnection flux tubes return inwards to the inner magnetosphere. Figure from \cite{delamere2013magnetotail}, based on \cite{vasyliunas_plasma_1983} and \cite{delamere_solar_2010}. }
    \label{fig:bag2014}
\end{figure}

Figure \ref{fig:bag2014} highlights structures within the Jovian magnetosphere. Closest to Jupiter there are Jupiter's rings (at 2-3 $R_J$) and intense radiation belts (up to 10 $R_J$) of (ultra) relativistic particles (detailed in the later section). At Io's orbit (5.9 $R_J$), the injections of iogenic plasma form the Io Plasma Torus (IPT) that encircles Jupiter (detailed in later section). From the IPT, over timescales of weeks, magnetic flux tubes interchange to transport iogenic plasma throughout the magnetosphere.  

\begin{figure}
    \centering
    \includegraphics[width=\textwidth]{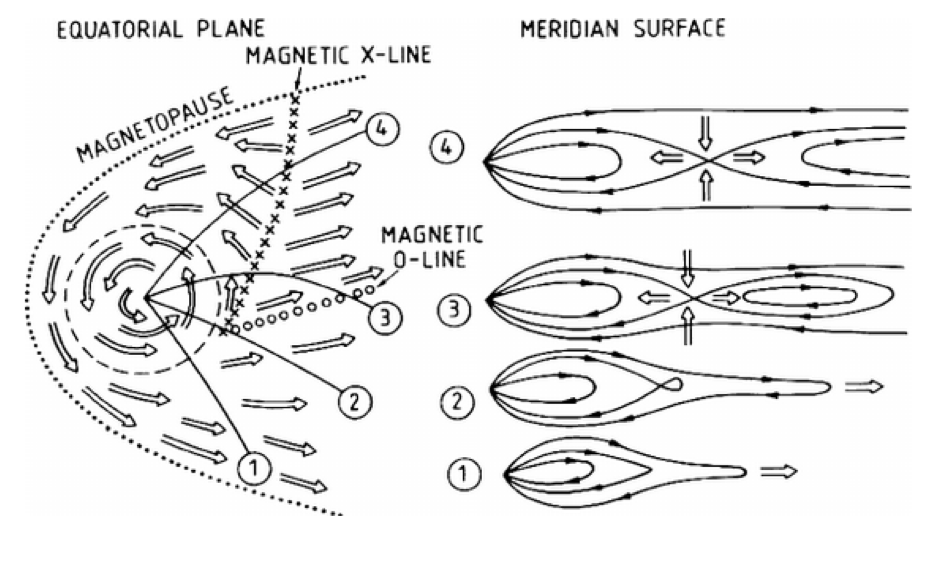}
    \caption{Schematic of Rotationally-Driven Tail Reconnection \cite{vasyliunas_plasma_1983}: The flow of magnetospheric plasma at Jupiter looking down onto the equatorial plane (left) and subsequent Jovian field configuration looking side-on in the meridianonal plane (right) with the stretching of field lines through tension from mass-loading and the subsequent release of a plasmoid (3 and 4) and dipolarisation of the magnetic field lines. Figure from \cite{vasyliunas_plasma_1983}.}
    \label{fig:vasyluinas1983}
\end{figure}

Jupiter's plasma is frozen into the magnetic field lines, which are anchored to the Jovian ionosphere and consequently rotate with the planet. This rapid rotation provides a centrifugal force that confines the plasma to the equatorial region, producing a `magnetodisk'/`plasmadisk'. As plasma rotates into the dusk and night-side tail, magnetic field lines become increasingly stretched by the equatorially confined plasma. Eventually, in the dusk to night-side tail the current sheet that separates the field lines becomes so thin that oppositely directed field lines can come into contact(see nightside blue field lines in Figure  \ref{fig:bag2014} for schematic). When this happens, tail-reconnection processes occur and the stretched field lines release plasmoids through the Vasyluinas Cycle \cite{vasyliunas_plasma_1983} (see Figure \ref{fig:vasyluinas1983} for details and red circles on the lower panel of Figure \ref{fig:bag2014}). This process of magnetic loading and unloading of the system is observed to happen on timescales of a few days \cite{vogt_reconnection_2010}.


Along the magnetopause boundary of the system, both reconnection processes with the solar wind and viscous processes such as Kelvin Helmholtz instabilities occur \cite{bunce_jovian_2004,delamere_solar_2010,masters_cassini_2010}. The relative contributions of each of these processes to the overall dynamics of the system remains a topic of debate \cite{cowley_modulation_2003, mccomas_jupiter_2007,cowley_comment_2008, delamere_solar_2010,masters_more_2018}.

Through these complex magnetospheric interactions, Jupiter produces a variety of distinct auroral emissions that include auroral footprints from its satellites \cite{Bhattacharyya_2018, Bonfond_2009, Bonfond_2013, Hue_2019, Szalay_2018}, diffuse and transient injection aurora \cite{wibisono_jupiters_2021,Kimura_2015,Mauk_2002, yao_reconnection-and_2020}, a semi-continuous oval of emission (`the main emission') \cite{Grodent_2003} and an array of vibrant and rapidly varying polar emissions \cite{Grodent_2015}. X-ray emissions connected with the satellite footprints are yet to be reported, but we discuss the rest of the X-ray auroral emissions in turn.

\subsection{Jupiter's Hard X-ray Aurorae}

\begin{figure}
    \centering
    \includegraphics[width=\textwidth]{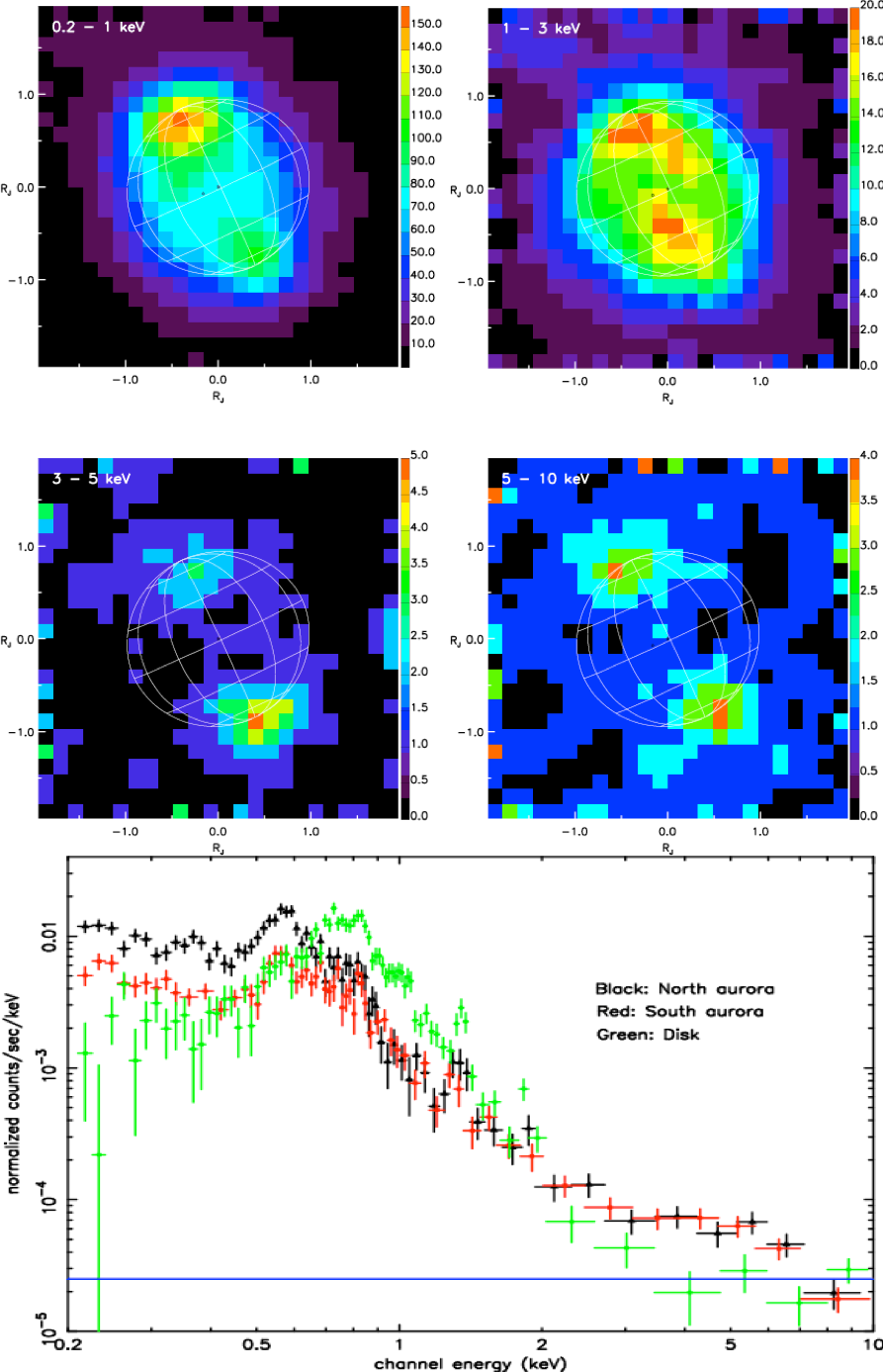}
       \caption{Top 4 panels: Smoothed \textit{XMM-Newton EPIC-pn} images of Jupiter from 0.2-1 keV (upper left), 1-3 keV (upper right), 3-5 keV (lower left) and 5-10 keV (lower right). The colour bar is in units of counts. A 30$^{\circ }$latitude-longitude graticule is overlaid. 
    Lower panel: Combined spectra from the equatorial region (green) and Northern (black) and Southern (red) X-ray aurorae, from observations in November 2003. The estimated EPIC particle background is shown by the horizontal blue line. Figures from \cite{branduardi-raymont_study_2007}.}
    \label{fig:gbrhard}
\end{figure}

While the expectations in the 1960s and 1970s were that Jupiter's X-ray emission would be dominated by hard X-ray bremsstrahlung from electrons, the first X-ray observations revealed that ion spectral lines actually dominated the emission. However, two decades later, the first \textit{XMM-Newton} observation of Jupiter did indeed show that alongside the spectral lines from precipitating ions, Jupiter's X-ray emission also included a bremsstrahlung component \cite{branduardi-raymont_first_2004,branduardi-raymont_study_2007} (Figure \ref{fig:gbrhard}). This component is sometimes best fit by a power-law and sometimes by a bremsstrahlung continuum, potentially indicative of non-thermal and thermal electron populations.

\begin{figure}
    \centering
    \includegraphics[width=\textwidth]{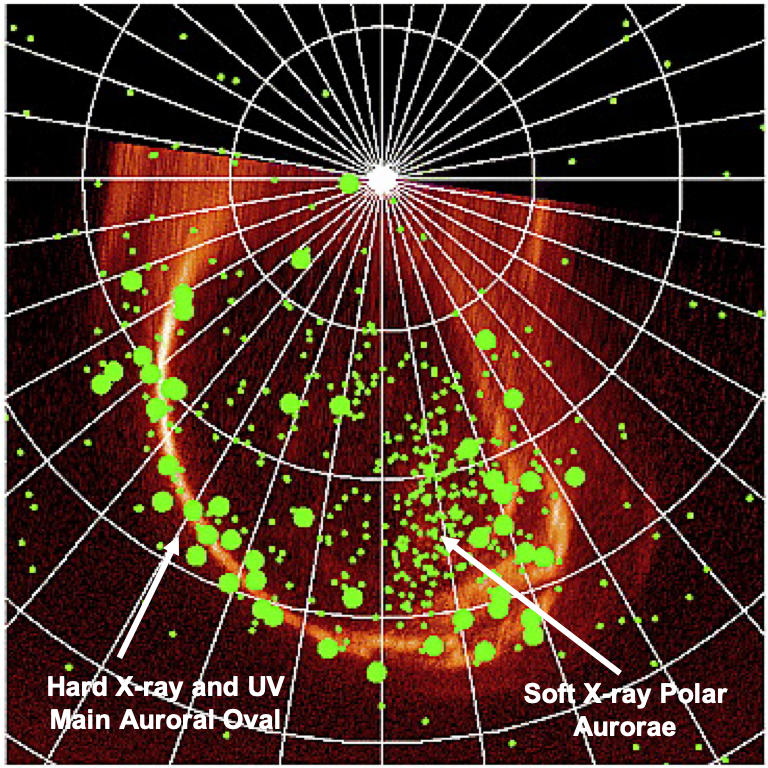}
    \caption{Overlaid projection of Jupiter's North pole showing the UV (orange) and X-ray (green dots) aurorae as recorded by the \textit{Hubble Space Telescope STIS FUV} instrument and \textit{Chandra ACIS} on 24 February 2003. Each green dot indicates an X-ray photon. Small green dots show $<$ 2 keV photons. Large green dots show $>$ 2 keV photons. The $10^\circ$ latitude-longitude spaced grid is fixed in System III with $180^\circ$ toward the bottom and $270^\circ$ to the left. Figure from \cite{branduardi-raymont_spectral_2008}.}
    \label{fig:gbr2008}
\end{figure}

\begin{figure}
    \centering
    \includegraphics[width=\textwidth]{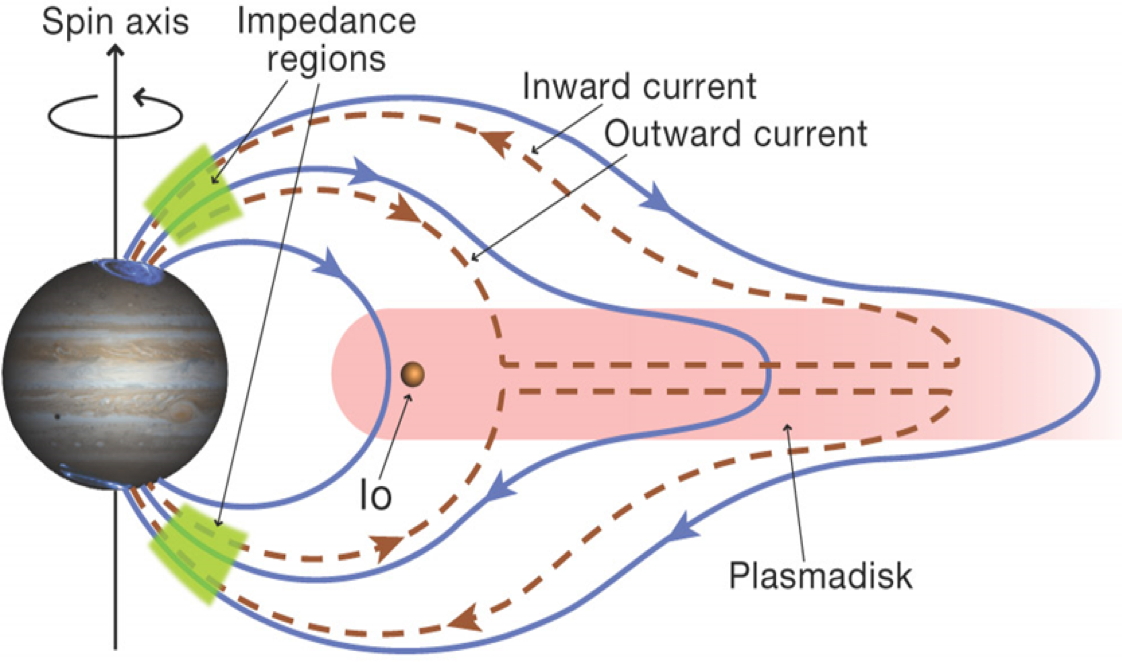}
    \caption{Schematic of Jupiter's outward and inward current system from
\cite{bagenal_magnetospheric_2014}. This shows the flow of an outward current system (brown dashed lines) anti-planetward, the subsequent radial flow which produces a J X B force (cross product of the current,  j, and magnetic field, B) to accelerate iogenic plasma in the plasmadisk that would otherwise be sub-corotating back towards corotation. To `complete the circuit' a subsequent inward current system returns the flow from the outer magnetosphere back to the ionosphere of the planet. The outward and inward currents flow along magnetic field lines (blue lines).}
    \label{fig:maukcurrents}
\end{figure}

Subsequent studies revealed that the hard X-ray emissions are not spatially co-located with the soft X-ray emissions. Through \textit{Chandra ACIS's} spatial and energy resolution it was found that the hard X-ray emissions (large green dots in Figure \ref{fig:gbr2008}) are co-located with the UV main auroral oval (orange oval-like structure in Figure \ref{fig:gbr2008}), suggesting that the same population of electrons that produce the UV main oval also produce the X-ray bremsstrahlung aurorae \cite{branduardi-raymont_spectral_2008}. 

The magnetic field lines that connect Jupiter's UV and hard X-ray main oval map to the middle magnetosphere between 15-50 $R_J$ from the planet. Here, the UV auroral footprint of the moon Ganymede provides a valuable reference point that enables us to identify the source location of the aurora \cite{grodent2009auroral}. It is thought that the electron precipitation that generates the UV main oval in this region is produced by a current system that transfers angular momentum from the planet to the surrounding magnetospheric plasma, enforcing the plasma's corotation with the planet \cite{cowley_origin_2001,hill_jovian_2001}. Figure \ref{fig:maukcurrents} shows a schematic of one theoretical version of the current system. The current flows outward from the planets pole, carried by precipitating electrons that produce the hard X-rays in the location of the UV main oval (Figure \ref{fig:gbr2008}). The current then flows radially out through the plasmadisk applying a j x B force (cross product of the current,j, and magnetic field, B) to the plasma, which acts to enforce co-rotation of the plasma. As the current reaches the outer magnetosphere, it is thought to return to the planet in the polar region, carried in-part by precipitating ions, associated with the soft X-rays (small green dots in Figure \ref{fig:gbr2008}) \cite{cravens_implications_2003}. For further details of the theoretical formulations of this current system see \cite{cowley_origin_2001,hill_jovian_2001}. 

Currently, Juno measurements are evolving our understanding of the particle precipitation and current systems in these regions. The results show that in actuality the UV and hard X-ray main oval has both inward and outward current systems interspersed within it \cite{kotsiaros2019birkeland}.  Stochastic processes in the magnetosphere (e.g. wave-particle interactions and turbulence) play a key role in the particle precipitation and therefore generating the auroral emissions \cite{mauk2017nature}. In the main oval, it seems that mono-directional electrons are accelerated by both electrostatic potentials and broadband processes, with the broadband processes dominating \cite{mauk2020energetic}. In contrast, in the region poleward of the main oval emission, bi-directional electrons are observed. These are predominately associated with broadband acceleration \cite{mauk2020energetic}. A detailed exploration of how the hard X-ray emissions connect to either broadband and/or electrostatic potentials is yet to be undertaken.

The UV main oval often brightens during compressions of the Jovian magnetosphere by the solar wind. In line with this, the hard X-ray emissions have also been observed to brighten during solar wind compressions \cite{dunn_impact_2016,dunn_comparisons_2020}. There also appears to be a correlation between brightening of Jupiter's main UV oval and loading and unloading of magnetic flux in Jupiter's tail\cite{yao_relation_2019}.  This process leads to intervals of brightening and dimming of Jupiter's UV main oval and thus is likely to modulate the X-ray bremsstrahlung emission over few-day timescales, although this is yet to be tested.

Figure \ref{fig:gbrhard} shows that the hard X-ray component from precipitating electrons is brighter from the Southern aurorae than the North. The soft X-ray aurorae show the opposite pattern and are brighter in the North than the South \cite{branduardi-raymont_study_2007}. One possible explanation for the brighter Southern hard X-ray emission  is that the South auroral oval is closer to the limb of the planet than the Northern oval. This means that from Earth, for the South, the precipitating electrons are moving at close to 90$^{\circ }$ to an observer, i.e. in the direction at which the bremsstrahlung emission is at its peak. In contrast, for the North, the electrons are precipitating closer to our line of sight, with bremsstrahlung produced at 90$^{\circ }$ to us. This effect becomes increasingly dominant for more relativistic electrons\cite{branduardi-raymont_study_2007}.

\begin{figure}
    \centering
    \includegraphics[width=0.7\textwidth]{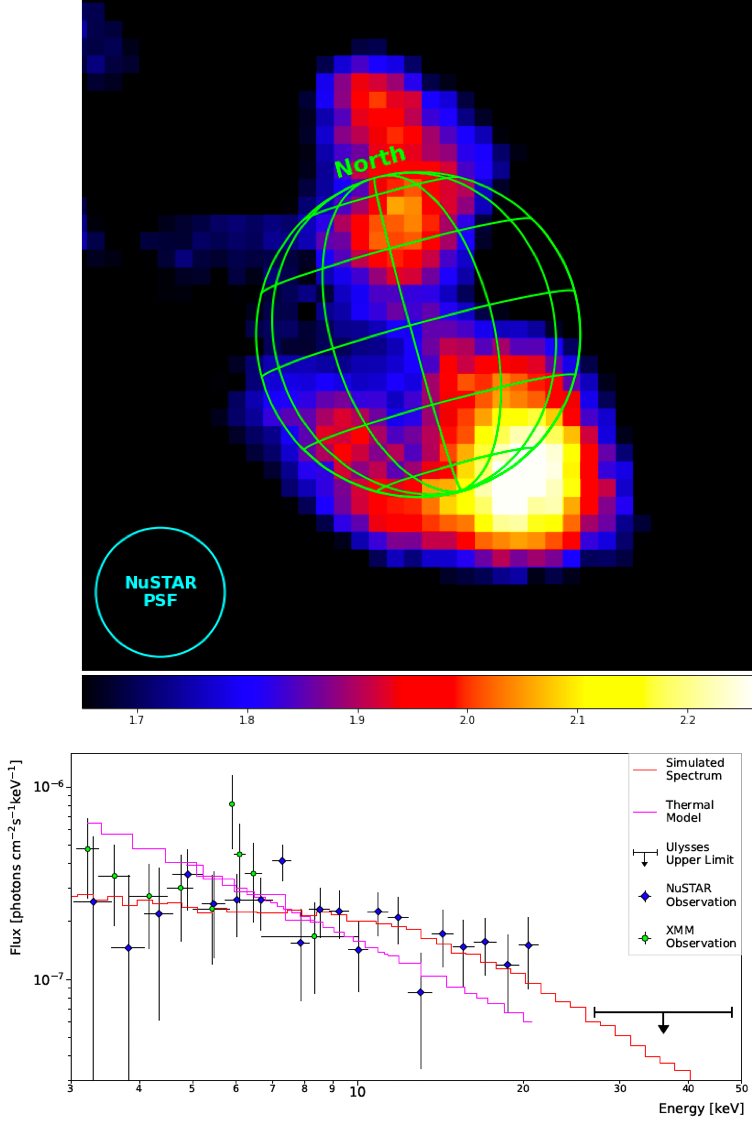}
    \caption{Upper Panel: \textit{NuSTAR} 8–20 keV image of Jupiter smoothed with a 6-pixel gaussian kernel and overlaid with a green graticule.
    Lower panel: \textit{XMM-Newton-EPIC} + \textit{NuSTAR} spectra of Jupiter (black) with simulated spectrum (red) and best-fit thermal bremsstrahlung model (kT = 200 keV; pink). The arrow indicates the 27–48 keV flux upper limits obtained by in-situ Ulysses measurements \cite{hurley_upper_1993}. Figures from \cite{Mori_NuSTAR}.}
    \label{fig:moriNuSTAR}
\end{figure}

\textit{XMM-Newton} has observed X-ray emissions from Jupiter up to 10 keV. Observations with the \textit{NuSTAR} telescope have shown that Jupiter produces hard X-ray emissions up to $\sim$20 keV and that these are predominantly produced from the Southern hemisphere with a flat power law of $\Gamma = 0.60 \pm 0.22$ (where X-ray photon flux $N(E) \sim E^{-\Gamma}$) indicative of non-thermal processes (Figure \ref{fig:moriNuSTAR} \cite{Mori_NuSTAR}). The GEANT4\cite{agostinelli_geant4simulation_2003} simulation toolkit (http://cern.ch/geant4) is commonly used to simulate the interaction of particles with matter and the production of X-rays. In this case, GEANT4 simulations, with inputs from in-situ Jupiter electron flux measurements by Juno, found that the electron population observed by Juno could produce the hard X-ray emissions observed by NuSTAR \cite{Mori_NuSTAR}.

\begin{figure}
    \centering
    \includegraphics[width=\textwidth]{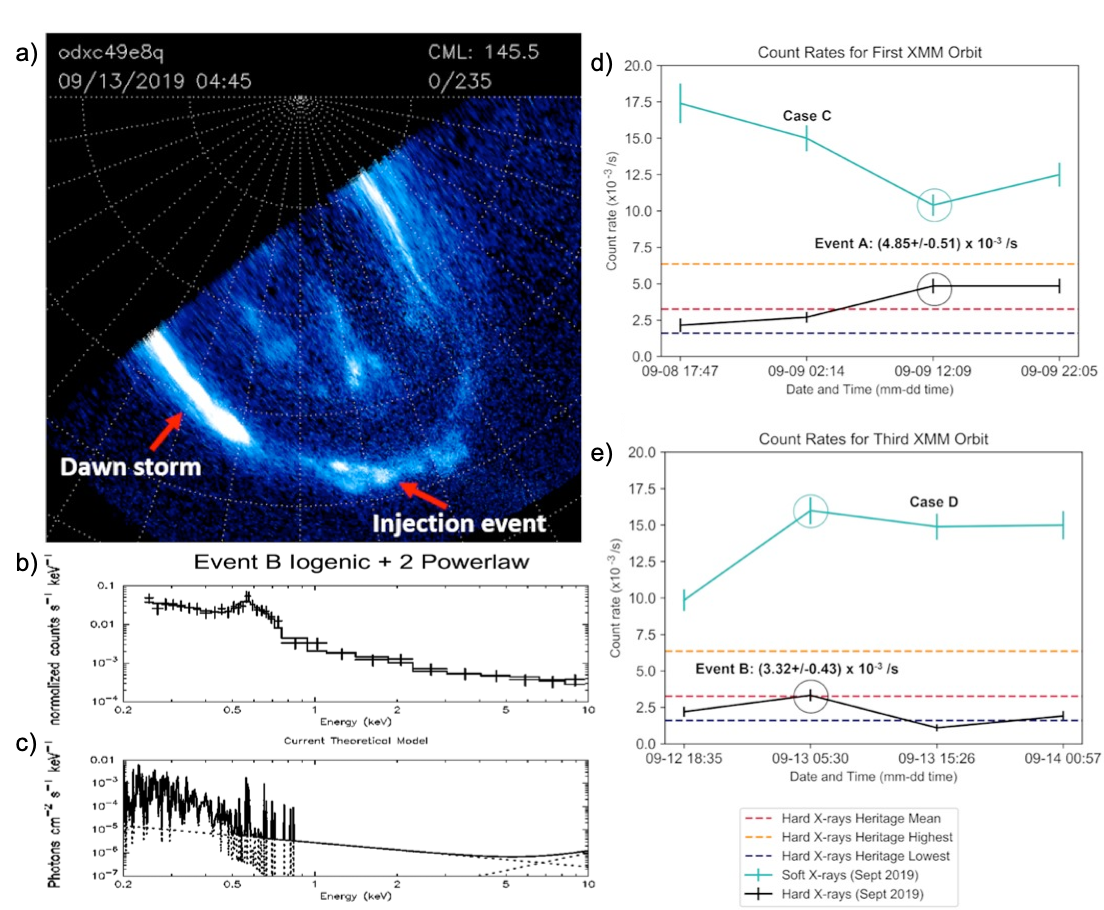}
    \caption{a) \textit{Hubble Space Telescope (HST)} polar projection of the FUV northern aurora showing a dawn storm and injection event in September 2019. b) \textit{XMM-Newton-EPIC-pn} spectrum from Jupiter's Northern aurorae (crosses) and best-fit model (black line) during the HST observation. c) Theoretical atomic charge exchange and power law models used to fit the spectrum. The solid lines display the dominant model and the dashed line the recessive model at a given energy. d) Count rates during a 4-Jupiter-rotation \textit{XMM–Newton} observation. The times are the mid-times (UTC) when the Northern X-ray aurora was in view. An interval when a dawn storm and injection event were present is highlighted by the circle. e): same as d for another \textit{XMM–Newton} observation. The aurora shown in a is highlighted by the circle in e). Figures from \cite{wibisono_jupiters_2021}.}
    \label{fig:wibisono2021}
\end{figure}

\subsubsection{Dawn Storms and Injections in the UV and Hard X-ray Aurorae}

Following observational hints that hard X-rays favour the dawn sector of the aurorae \cite{dunn_impact_2016, dunn_jupiters_2020}, recent studies have shown correlations between hard X-ray emissions and transient brightenings of the UV auroral oval referred to as `dawn storms' (see Figure \ref{fig:wibisono2021}a). UV observations show that dawn storms are often observed at the same time as auroral injections (Figure \ref{fig:wibisono2021}a shows these two types of auroral emissions) \cite{wibisono_jupiters_2021}. Comparisons with \textit{Juno} in-situ data showed that dawn storms coincide with magnetic reconnection in the tail region of Jupiter's magnetosphere\cite{yao_reconnection-and_2020}. These isolated and localised auroral brightenings typically only last one Jupiter rotation, and therefore are not consistent with solar wind compressions, which cause a global brightening of the main oval for several Jupiter rotations. Instead, dawn storms are thought to be caused by intervals of internally driven magnetic reconnection associated with the Vasyluinas cycle \cite{vasyliunas_plasma_1983}. As discussed previously, these occur naturally in the Jupiter magnetosphere, as Io's volcanoes inject mass into the system and this mass-loading applies tension to the magnetic field lines in the tail, which stretch until they reconnect to release mass from the system as a plasmoid (see Figure \ref{fig:vasyluinas1983} for schematic). Having released this mass, the field lines will return from their stretched state to a more dipolar state, and this dipolarisation of the magnetic field lines is thought to produce the currents and flows of electrons that generate the auroral injection signatures \cite{yao_reconnection-and_2020}. Figure \ref{fig:wibisono2021}d and e) show that Jupiter's hard X-ray emissions also respond to this chain of physical processes in the Jovian magnetosphere, with hard X-ray emission brightening at the same time as dawn storm and injection aurorae are seen, caused by magnetic reconnection and dipolarisation \cite{wibisono_jupiters_2021}. Furthermore, there is some suggestion of multiple power law continua (Figure \ref{fig:wibisono2021}b and c)\cite{wibisono_jupiters_2021} that may each respectively connect to the dawn storm and injection aurora structures.

Figure \ref{fig:wibisono2021}d) and e) highlight a further curious property of Jupiter's aurorae: the independent behaviours of Jupiter's hard and soft X-ray emissions. While these emissions sometimes brighten together (Figure \ref{fig:wibisono2021}e), they can also brighten independently of one another (Figure \ref{fig:wibisono2021}d). We now turn to this seemingly independent soft X-ray aurora.

\subsection{Jupiter's Polar Soft X-ray Aurorae}

While Jupiter's most energetic photons come from the planet's hard X-ray auroral oval, the dominant contribution in photon number are the soft X-ray line emissions from the polar aurorae (Figure \ref{fig:gbr2008}). These can dominate the entire planet's luminosity. We first look at the X-ray production mechanism for these lines before turning to the auroral structures and the processes that lead to the particle precipitation.

The soft X-ray auroral spectrum (e.g. Fig. \ref{fig:gbr2004xmm} and \ref{fig:gbrhard}) is produced by precipitating ions. Upon colliding with Jupiter's atmosphere, these ions undergo charge exchange interactions with the neutral atmosphere and produce an array of spectral line emissions \cite{cravens_auroral_1995,branduardi-raymont_first_2004,elsner_simultaneous_2005,kharchenko_ion_2006,branduardi-raymont_study_2007,kharchenko_modeling_2008,hui_ion-induced_2009,hui_comparative_2010,dunn_impact_2016,dunn_jupiters_2020,dunn_comparisons_2020} (see Fig. \ref{fig:gbr2004xmm}). Oxygen charge exchange X-ray lines are a common feature of Jupiter's auroral spectrum, such as those from this example transition:

\begin{equation}
    O^{8+} + H_2 \rightarrow O^{7+} + H_2^+ + \textrm{X-ray}
\end{equation}

Spectral fits for Jupiter's aurorae most commonly involve a combination of sulphur and oxygen spectral lines, representative of Iogenic plasma precipitating from the magnetosphere. However, for some observations solar wind ions need to be included in the models to produce the observed charge exchange spectra \cite{elsner_simultaneous_2005,branduardi-raymont_first_2004,hui_comparative_2010, dunn_jupiters_2020,wibisono_temporal_2020}. It remains an open question whether the solar wind ions precipitate on field lines open to the solar wind or whether these ions enter the outer magnetosphere through e.g. Kelvin Helmholtz instabilities and then precipitate as part of the magnetospheric plasma \cite{dunn_comparisons_2020}.

\begin{figure}
    \centering
    \includegraphics[width=0.7\textwidth]{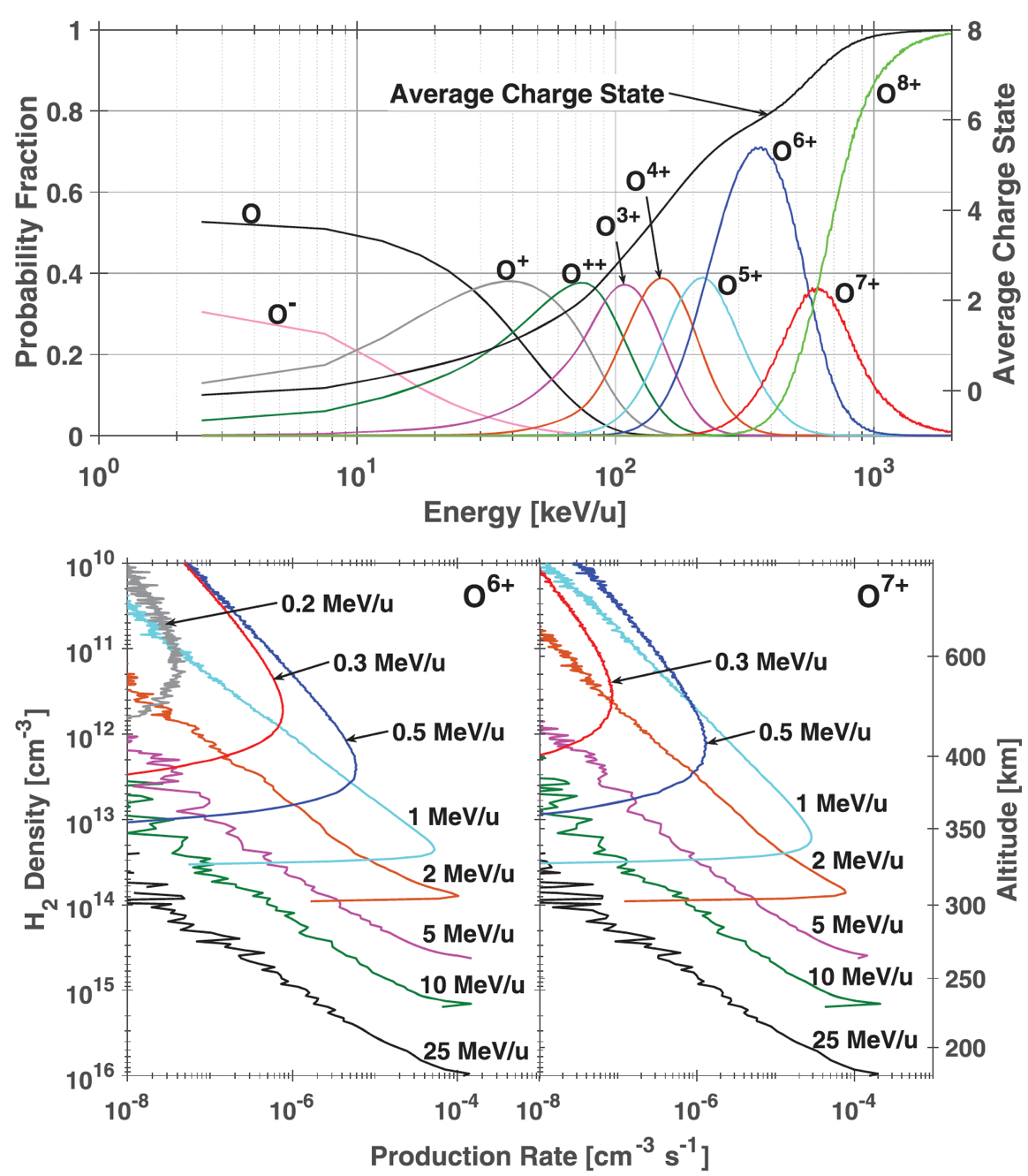}
    \caption{Upper panel: Resulting Oxygen charge state distribution as a function of ion energy, upon precipitating into the Jovian atmosphere. Previous work \cite{houston_jovian_2018, ozak_auroral_2010} had the $O^{6+}$ peak at $\sim$900 keV/u. Newly developed cross sections, lead the peak to be shifted to an energy of $\sim$350 keV/u. Lower panels: The $O^{6+}$ (left) and $O^{7+}$ (right) production rates vs $H_2$ density and altitude for the different incident ion energies shown. Production rates are normalized to 1 incident ion $cm^{-2} s^{-1}$ \cite{houston_jovian_2020}. Figures from \cite{houston_jovian_2020}.}    
    \label{fig:houston2020}
\end{figure}

The solar wind already has high abundances of high charge states of oxygen needed to produce the observed charge exchange emissions (i.e. $O^{7+}, O^{8+}$), but such high charge states are less abundant in the Jovian magnetosphere, where charge states such as $O^{+}$ and $O^{2+}$ are most common. Consequently, additional processes to strip charges from a precipitating ion are required in order to produce the observed X-ray signatures. Upon precipitating into the atmosphere, an oxygen ion with an energy of greater than 100 keV/amu has the possibility to strip enough charge to produce an X-ray through subsequent charge exchange (see Figure \ref{fig:houston2020}). An example single ionisation - single charge stripping interaction is:
\begin{equation}
O^{+}+H_2 \rightarrow O^{2+}+H_2^+ + 2e^-
\end{equation}
A huge range of interactions that redistribute energy and/or charge are possible for an ion colliding with a neutral, including: single/double ionisation, single/double projectile excitation, single/double charge stripping, charge exchange ionisation, double capture autoionisation, single/double electron capture and excitation. Comprehensive lists of interactions can be found in the wealth of Monte Carlo modelling work on ion precipitation in Jupiter's aurorae \cite{kharchenko_ion_2006,kharchenko_modeling_2008,ozak_auroral_2010,ozak_auroral_2013,houston_jovian_2018,houston_jovian_2020}. Notably, recent work found that the energy required to produce sufficiently high charge state ions is a factor of 3 lower than previously thought \cite{houston_jovian_2020}, meaning that less extreme accelerations are required to generate Jupiter's observed soft X-ray aurorae than was thought before 2020. Once an ion has reached a sufficiently high charge state, and has shed enough energy, it can undergo charge exchange and recombination interactions that produce the observed soft X-ray auroral emissions\cite{houston_jovian_2020}. 

\begin{figure}
    \centering
    \includegraphics[width=0.65\textwidth]{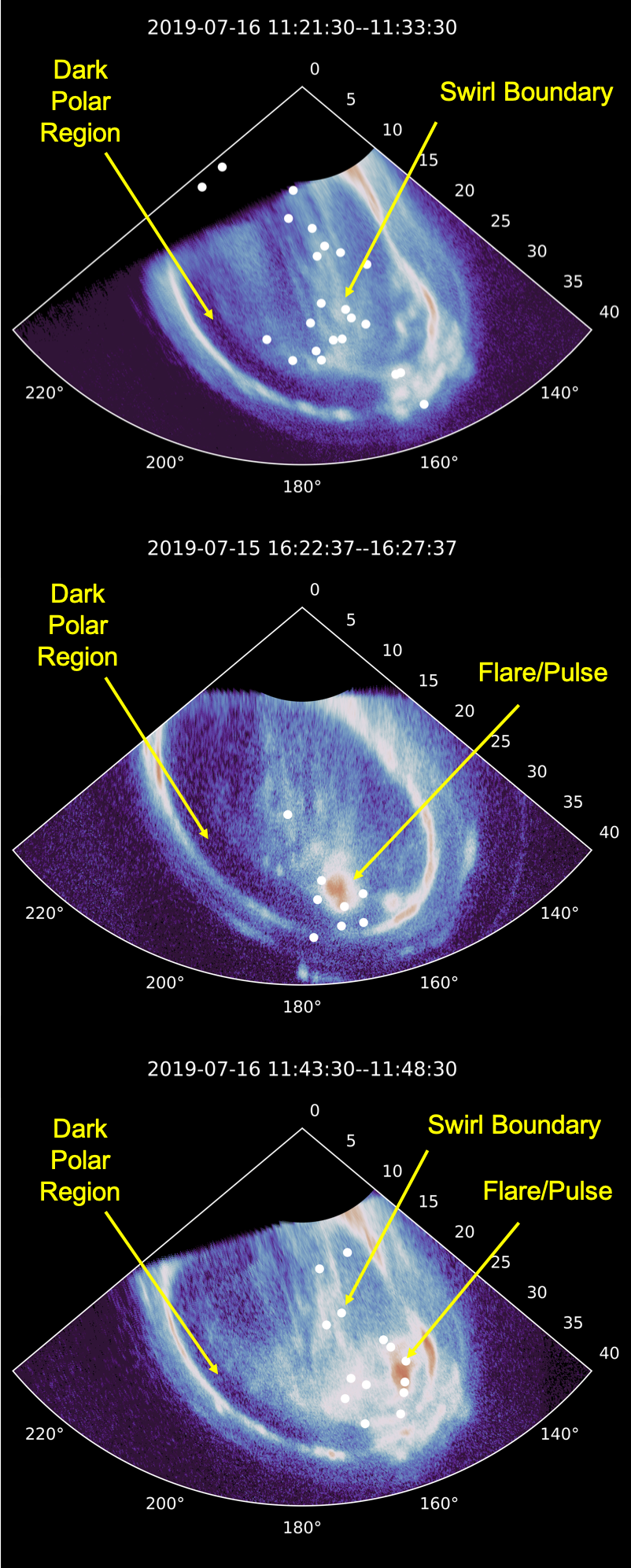}
    \caption{North Pole Projections of Jupiter's UV and X-ray Aurora from \textit{Hubble Space Telescope (HST)} and \textit{Chandra} X-ray  observations in July 2019. UV brightness is the mean across the interval. X-ray photons (white dots) are all those collected during the interval. Top: 16 July 11:21-11:31 shows more diffuse/patchy UV and X-ray aurora on the boundary of the `swirl region'.  The longer integration time of the observation enables this connection to be observed. Middle: 15 July 16:22-16:26 shows a UV and X-ray flare (sometimes called an X-ray 'pulse') in the `active region'. Bottom: 16 July 11:43-11:47 shows an example of simultaneous emission from both the swirl boundary and a flare/pulse in the active region. These two emissions occur independently, with the swirl boundary emission having a steady/flickering time signature \cite{dunn_comparisons_2020} and the flares/pulses consisting of an impulsive short-lived brightening \cite{gladstone_pulsating_2002,dunn_independent_2017}. All three panels show that no X-ray photons are detected from the UV Dark Polar Region. Longitude is indicated along the arc of the projection, with latitude along the radius; both are in degrees. The UV colour bar saturates in red at 1 MR. Figure based on \cite{dunn_DPR}.}
    \label{fig:swirlandflares}
\end{figure}

Early \textit{Chandra} observations of Jupiter suggested that the polar aurora was concentrated into a `hot spot' that rotated with the planet \cite{gladstone_pulsating_2002}. This `hot spot' can be seen through the clustering of soft X-ray photons (small green dots) seen in Figure \ref{fig:gbr2008}. Subsequent work showed that there was a concentration of photons in this region and also an extended transient `halo' of emission surrounding the hot spot\cite{kimura_jupiters_2016}. These emissions are caused by processes in Jupiter's outer magnetosphere beyond 50 $R_J$ and may relate to processes at the magnetopause \cite{weigt_chandra_2020,weigt_characteristics_2021,vogt_improved_2011,vogt_magnetosphere-ionosphere_2015}. Recently, it has been shown that the soft X-ray `hot spot' is not a single coherent structure but instead two distinct auroral structures, which may each be produced by very different processes: flares (sometimes called pulses) of emission in the active region \cite{gladstone_pulsating_2002, elsner_simultaneous_2005, dunn_independent_2017} and a dimmer but more continuous emission from the boundary of the swirl region (see Figure \ref{fig:swirlandflares}). The swirl region is the auroral region closest to Jupiter's magnetic pole and is named for the intermittent emissions that often form swirling morphology \cite{masters_magnetic_2021}. We discuss each in turn.

\subsubsection{Pulsed X-ray Ion Auroral Flares/Pulses}

\begin{figure}
    \centering
    \includegraphics[width=0.85\textwidth]{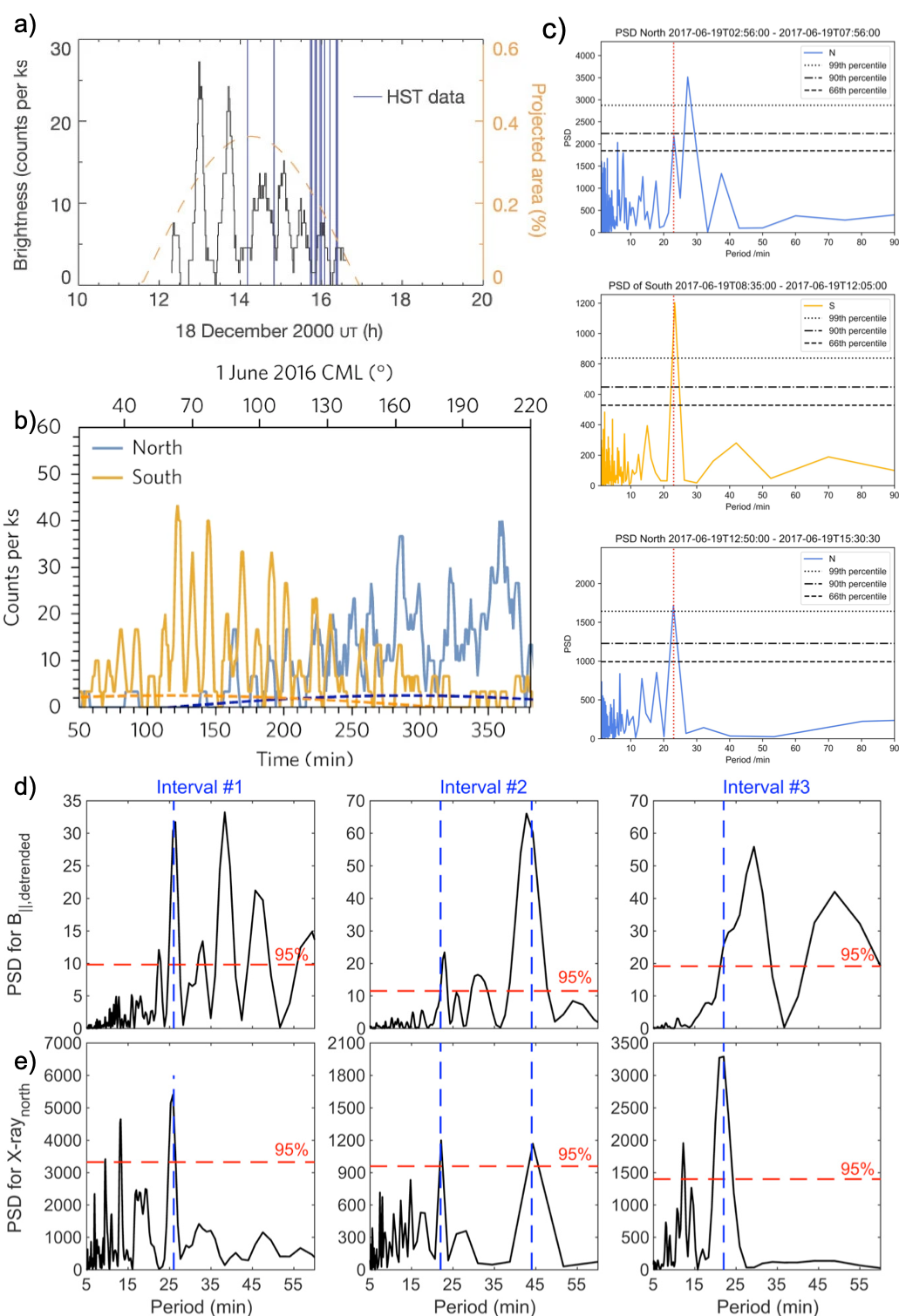}
    \caption{a) \textit{Chandra HRC} lightcurve from Jupiter's Northern X-ray `hot spot' on 18 December 2000, with an 11-min boxcar smoothing of 1-min binned data. The dashed orange line shows, as a percentage of the projected area of Jupiter, the projected area of the hot spot. A simultaneous \textit{HST-STIS} observation was obtained during the time shown. During this observation, the X-ray aurorae pulse with a 45-minute period. b) Chandra X-ray lightcurves from an interval when the northern (blue) and southern (gold) aurora were both observable on 1 June 2016. The visibility as a fraction of maximum visibility for the northern (blue) and southern (gold) hot spots is indicated by the dashed curves. The Central Meridian Longitude (CML) of Jupiter is shown across the top, and minutes from the observation start time (11:32 UT) are indicated on the bottom x-axis. The lightcurves are one-minute binned, with six-minute moving-average smoothing. c) Power Spectral Density Plots from Fast Fourier transformations of \textit{XMM-Newton EPIC-pn + MOS1 + MOS2} lightcurves from Jupiter's Northern (blue) and Southern (yellow) aurorae in the order (from the top) that each aurora viewing occurred. The 66th, 90th, and 99th percentiles are marked by the dashed, dashed dot, and dotted black lines, respectively. The vertical red dashed line marks the recurring period at $\sim$23 min. d) Lomb-Scargle periodograms of the detrended field-aligned magnetic field $B_\parallel$ (upper 3 panels) and e) northern X-ray emission (lower 3 panels) for three intervals when the Northern aurora was in view, during a 26 hour observation. Each interval lasted 3-5 hours. The red dashed lines indicate the confidence level of 95\% for each analysis. The blue dashed vertical lines show shared x-ray and magnetic field power spectral density (PSD) periodogram peaks.  Figures from a) \cite{gladstone_pulsating_2002}, b) \cite{dunn_independent_2017}, c) \cite{wibisono_temporal_2020}, d) \cite{yao_revealing_2021-2} }
    \label{fig:pulsations}
\end{figure}

The first \textit{Chandra} observations of Jupiter revealed perhaps the most puzzling property of the soft X-ray emissions. The X-ray flux from this region is not produced by a steady precipitation of ions, but instead is highly pulsed, with impulsive bursts of emission occurring on timescales of minutes to tens of minutes \cite{gladstone_pulsating_2002,jackman_assessing_2018,dunn_independent_2017}. Intriguingly, sometimes this emission is observed to have a regular beat. Figure \ref{fig:pulsations} shows examples of intervals with a) 45-minute pulsations from the Northern aurorae\cite{gladstone_pulsating_2002}, \ref{fig:pulsations}b) 9-13 minute pulsations from the South (sometimes independent of Northern emission) \cite{dunn_independent_2017} and \ref{fig:pulsations}c) a shared 23-minute Northern and Southern periodicity lasting for a Jupiter rotation \cite{wibisono_temporal_2020}. 
These impulsive bursts of X-rays have been seen to occur simultaneous with UV flares that can reach teraWatt powers \cite{elsner_simultaneous_2005, waite_auroral_2001} (such UV flares can be seen in the middle and lower panels of Figure \ref{fig:swirlandflares}).

\begin{figure}
    \centering
    \includegraphics[width=\textwidth]{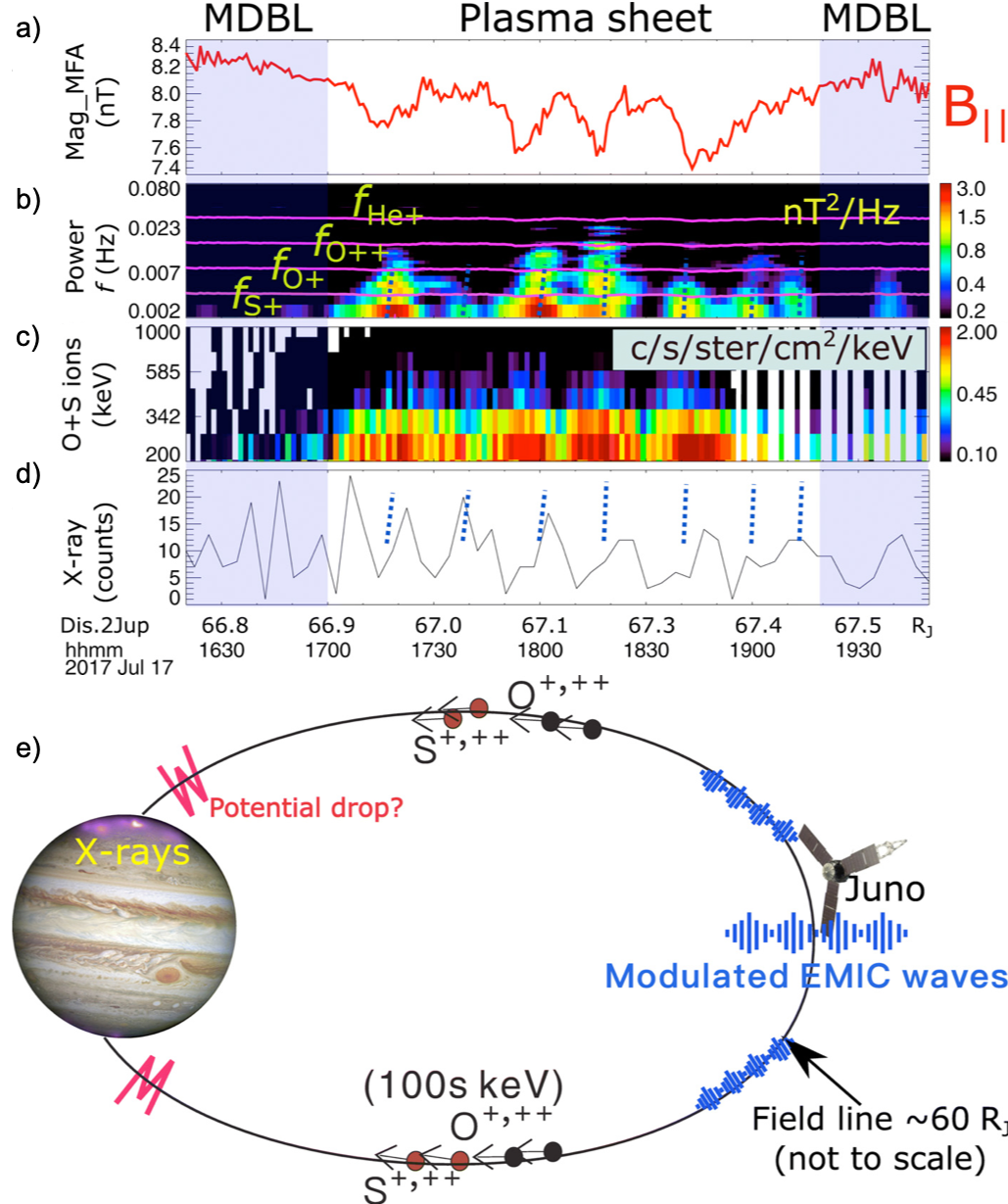}
    \caption{Comparison of \textit{Juno} and \textit{XMM-Newton} measurements: a) \textit{Juno MAG} measurements of the field-aligned magnetic component ($B_\parallel$) in mean field–aligned coordinates (coordinates obtained over a 60-min window), showing compressional waves as dips in the magnetic field strength. b) Power spectral density of the magnetic field perturbations with the gyro frequencies of various charge states of ions ($He^+, O^{++}, O^+$, and $S^+$) overlaid, showing Electromagnetic Ion Cyclotron (EMIC) waves synchronised with the compressional waves. c) \textit{Juno JEDI} instrument measurements of the sulfur and oxygen energy and intensity, showing pulsations in the ion intensity in phase with the compressional waves and EMIC waves. d) XMM EPIC-pn and MOS light curves (binned to a resolution of 4 min) from the Northern X-ray aurora, showing X-ray pulses with the same pulation-rate as the compressional waves, EMIC waves and ion pulsations. The X-ray light curves have been shifted to account for the light travel time from Jupiter to \textit{XMM-Newton}. Blue dashed lines show the times of EMIC waves. The time taken for ions to precipitate along the magnetic field lines from the outer magnetosphere is expected to be minutes to tens of minutes so that it is not yet possible to directly connect a single EMIC wave with a single x-ray pulse. e) A schematic to illustrate the wave-particle interaction and the consequent x-ray emissions. Figures from \cite{yao_revealing_2021-2}.}
    \label{fig:emic}
\end{figure}

Studies leveraging the simultaneous \textit{Juno} and \textit{XMM-Newton} observations have revealed the chain of processes responsible for these pulsed X-ray emissions \cite{yao_revealing_2021-2}. The observations showed that Jupiter's soft X-ray aurorae pulsed in time with fluctuations in Jupiter's magnetic field strength that occur in the compressional direction for the field. Figure \ref{fig:emic}a) shows an example of these compressional slow-mode waves - periodic dips in the magnetic field. When these compressional slow-mode waves changed pulsation rate, the aurorae responded in kind (e.g. Figure \ref{fig:pulsations}d and e).

Further, the observations revealed the processes through which the potentially global-scale compressional waves impart their pulsations to the small-scales of the ion population. The large-scale compressional waves trigger EMIC waves (electromagnetic ion cyclotron waves), which operate on smaller scales, as resonances between the magnetic field and the ion gyration\cite{yao_revealing_2021-2}.  Figure \ref{fig:emic}b) shows the presence of the EMIC waves - indicated by the wave power cut-off at the gyro-frequencies of the heavy ions in Jupiter's magnetosphere (purple lines). The EMIC waves pitch angle scatter Jupiter's energetic ions, shepherding them along the magnetic field lines (see schematic in Figure \ref{fig:emic}e). Eventually these pulsations of ions reach the pole of the planet, colliding with the atmosphere to produce the charge exchange emissions observed (Figure \ref{fig:emic}d). For \textit{Juno} to detect the compressional waves and EMIC waves in the X-ray source region, it needed to be located $\sim$ 60$R_J$ from the planet \cite{dunn_impact_2016,kimura_jupiters_2016,weigt_chandra_2020,weigt_characteristics_2021}, and so consequently there is a consistent time-lag between each wave observation and the resulting auroral pulsation. 

While the EMIC waves scatter ions to direct them towards the pole, acceleration of the ions is also needed. MegaVolt potential drops were predicted from theory and subsequently observed in the polar regions of Jupiter's aurorae and provide the needed acceleration \cite{cravens_implications_2003,clark_heavy_2020}.

Given the vast differences in size and energy scale between the Jovian and terrestrial systems, it is perhaps surprising that EMIC waves govern ion precipitation to form aurora at both planets (proton aurora at Earth and CX aurora at Jupiter). This may hint at a universality in the role of compressional waves and EMIC waves in transporting particles and energy in space plasmas, independent of scale and energy. However, it remains unclear why the emissions are quasi-periodic. One possible explanation is that the compressional waves are reflected at cavities in the system \cite{kivelson_global_1984} e.g. between the magnetopause and the strong density gradient at the Io Plasma Torus. If that is the case, then the periodicity may be indicative of the scale of the system.  Furthermore, quasi-periodic pulsations are not unique to the X-rays, but are also observed in other wavebands (e.g. kHz radio emissions). How the multi-waveband periodic emissions connect remains to be explored.

It is also unclear why the soft X-ray aurora sometimes pulses and sometimes does not - i.e. what `switches on' the compressional waves. Several studies have shown that the auroral pulsations appear to be triggered by solar wind compressions \cite{dunn_impact_2016,wibisono_temporal_2020,weigt_chandra_2020, dunn_comparisons_2020}, which would disturb the system and trigger the compressional waves. Conversely, the \textit{XMM-Newton} observations during dawn storms and injection aurorae did not detect soft X-ray pulsations, so it may be that the (re)configuration of the magnetic field at these times inhibits the conditions necessary for compressional waves \cite{wibisono_jupiters_2021}. Systematic, statistical studies need to be undertaken to explore the causes.

\subsubsection{Swirl/Flickering Polar Soft X-ray Aurora}

Studies of Jupiter's Northern aurorae showed that X-ray pulsations were not the only soft X-ray auroral emission; a dim semi-continuous flickering emission is also sometimes present \cite{dunn_comparisons_2020}. Integrating observations over 10s of minute timescales reveals that these emissions (1-2 counts per minute) form arc structures coincident with bright UV emission along the boundary of the swirl region (e.g. Figure \ref{fig:swirlandflares}a and c). Observations suggest that when the flickering emissions are present, spectral models require the inclusion of solar wind ions \cite{dunn_comparisons_2020}. 

While the flared/pulsed emissions are strongly correlated with compressional and EMIC waves, several other processes have previously been proposed to explain Jupiter's soft X-ray aurora and may be relevant for the swirl boundary emission. We briefly explore these here. Figure \ref{fig:maukcurrents} shows that the outward currents that generate the main oval must return to the planet as inward/return currents. These return currents are carried by outward flowing electrons and precipitating ions - the signature of which may be the soft X-ray aurora \cite{cravens_implications_2003}. Many studies have suggested that Jupiter's aurorae connect to a region close to the magnetopause \cite{dunn_impact_2016, kimura_jupiters_2016, weigt_chandra_2020, weigt_characteristics_2021} and may therefore indicate dayside reconnection \cite{bunce_jovian_2004} or Kelvin Helmholtz instabilities \cite{dunn_independent_2017, kimura_jupiters_2016}, along this boundary. The solar wind ion signatures associated with the flickering emission \cite{dunn_comparisons_2020} may well provide a valuable clue worthy of systematic exploration. 

Alternatively, at such high latitudes, the Alfv\`{e}n travel times from the pole of the planet become significant fractions of the planet's rotation rate, meaning that magnetic field lines can begin to become twisted \cite{zhang_how_2021}. This twisting enables reconnection close to the planet with large voltages ($1 V m^{-1}$) similar to the solar corona \cite{masters_magnetic_2021}. The energetic particle injections caused by such processes may also generate X-ray emissions.

\subsubsection{Jupiter's Dark Polar Region}

The auroral studies discussed hitherto concentrated on the presence of auroral emissions, but the absence of emissions can also be indicative of the underlying physics in a magnetosphere. In the dawn sector, between the main oval and the swirl region there is a region mostly devoid of UV emission: the dark polar region (DPR) \cite{swithenbank-harris_jupiters_2019}. Figure \ref{fig:swirlandflares} shows that this region also appears to be empty of X-ray emission. Across the 14 simultaneous \textit{HST} and \textit{Chandra} Northern aurorae observations, the X-ray emission from the DPR was consistent with 0 counts \cite{dunn_DPR}. 

The Earth has a large dark region poleward of its main auroral oval, due to a significant proportion of the region being open magnetic field lines. In contrast, Jupiter's polar region is very dynamic with a variety of different auroral processes generating emission. The extent to which Jupiter is open to the solar wind or a completely closed system remains unknown. Perhaps this relatively small, crescent-like dark region is Jupiter's open field line region \cite{dunn_DPR,cowley_jupiters_2003,zhang_how_2021}.

\section{\textit{Direct Imaging of Jupiter's Surrounding Space Plasma}}

A powerful advantage of X-ray astronomy is the ability to directly image large regions of plasma. For planetary science and space science, where measurements are typically made of individual locations at single instances of time, these global view points are invaluable. For Jupiter, two regions so far have been imaged by X-ray observatories: the Io Plasma Torus and the Radiation Belts. Both structures would benefit from extensive additional exploration of the archive of X-ray observations. Additionally, X-ray emissions might be produced by charge exchange between Jupiter's magnetosheath and exosphere and between the magnetospheric plasma and the exosphere. Although, these are yet to be explored in analytical, modelling or observational work. This section will focus on the two regions that have been observed to date.

\subsection{X-rays from the Io Plasma Torus}

Jupiter's moon, Io, is the most volcanic body in the solar system. Every second, it ejects $\sim$1 ton of volcanic material into the space environment around Jupiter. Approximately 250 kg/s of the ejected neutral material (predominantly sulphur and oxygen) becomes ionised by solar photons or through collisions, and contributes to the dense toroidal cloud of plasma that encircles Jupiter between 5 to 10 $R_J$ as the Io Plasma Torus \cite{bagenal_empirical_1994,phipps_two_2021}. This region acts as the source for the majority of the plasma in the Jovian magnetosphere and so a detailed characterisation of its structures and dynamics at a global scale is key to understanding the jovian magnetosphere.

\begin{figure}
    \centering
    \includegraphics[width=\textwidth]{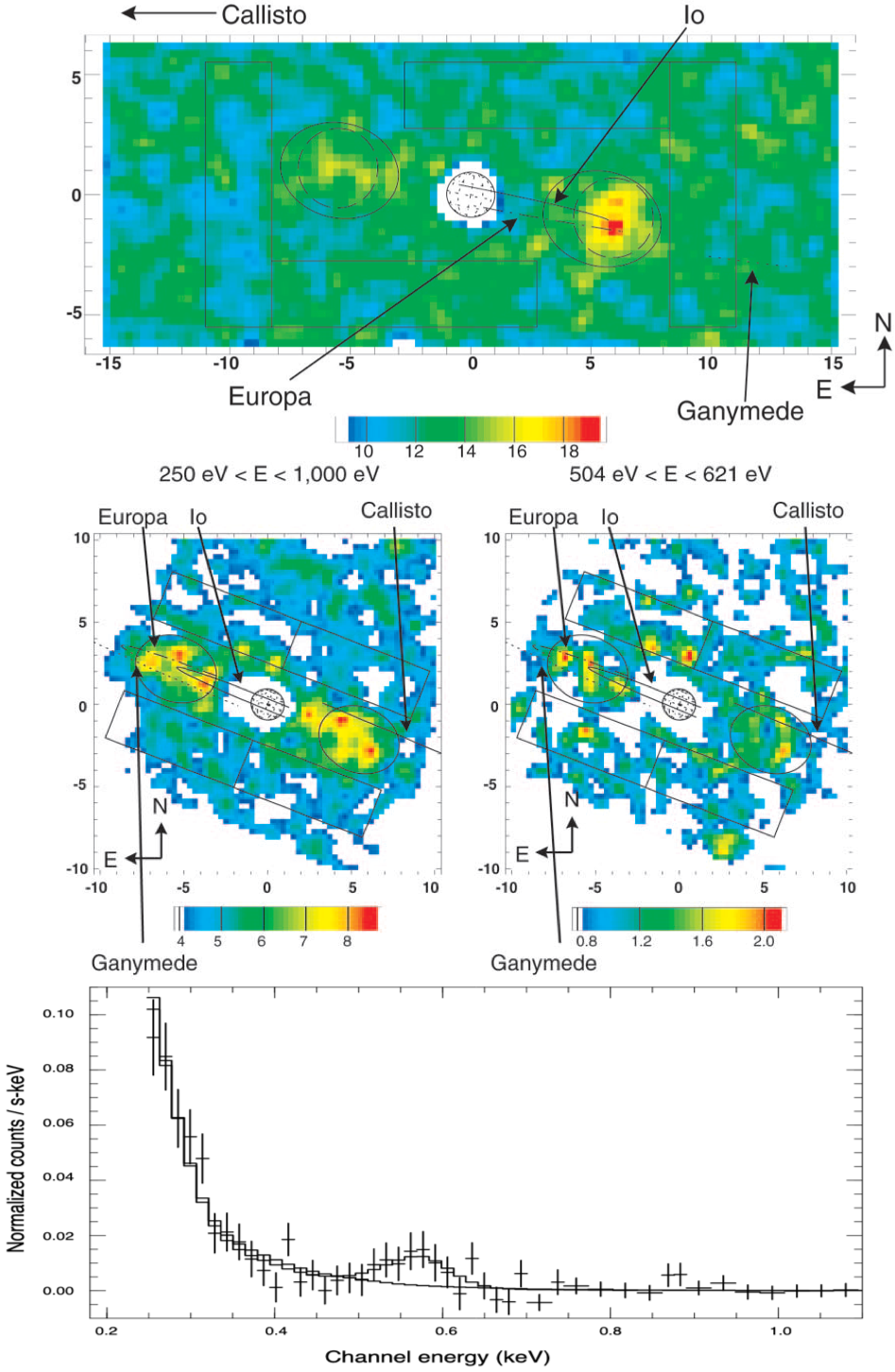}
    \caption{Upper panel: Gaussian ($\sigma = 7''.38$) smoothed \textit{Chandra HRC-I} image of the Io Plasma Torus (2000 December 18). Axes are in units of $R_J$, and the scale bar is in smoothed counts per image pixel. The paths traced by the moons are shown on the figure for Io (solid line), Europa (dashed line), and Ganymede (dotted line). Middle panel: As above, but for \textit{Chandra ACIS} image of the IPT (1999 November 25–26) for $250 eV <E <1000 eV$ (left) and for $504 <E <621 eV$ (center) and with Callisto's path (solid line on the dusk side). Lower panel: Background-subtracted \textit{Chandra ACIS} IPT spectrum, with a power-law model fit and also a power-law plus Gaussian model fit. Figures from \cite{elsner_discovery_2002}. }
    \label{fig:elsner2002ipt}
\end{figure}

\textit{Chandra ACIS} and \textit{HRC-I} observations of Jupiter, in November 1999 and December 2000, conducted the first direct images of X-ray emission from the Io Plasma Torus\cite{elsner_discovery_2002}. The top panel of Figure \ref{fig:elsner2002ipt} shows the X-ray image taken in December 2000 by the \textit{Chandra HRC-I}. The dusk side of the IPT (pointing West) appears brighter than the dawn side at this time - reminiscent of EUVE observations \cite{gladstone_recent_1998}. The \textit{Chandra ACIS} observations (Figure \ref{fig:elsner2002ipt} middle panel) show variation in the spatial distribution of the X-rays with energy, with a dawn enhancement in emission at energies associated with O VII lines (504-621 eV). Interestingly, this occurs at a time when Io and Europa were in the dawn sector. In contrast, the 250-1000 eV emission is roughly evenly distributed between the dawn and dusk sides of the IPT. This analysis of the X-ray IPT, recorded fluxes of 2.4 x $10^{14}$ ergs s$^{-1}$ cm$^2$, corresponding to a luminosity of 0.12 GW. These fluxes are $\sim$3 orders of magnitude below that reported from EUVE observations of the IPT \cite{gladstone_recent_1998}.

Analysis of the spectrum (Figure \ref{fig:elsner2002ipt} bottom panel) showed that best fit models used a power law or bremsstrahlung component to capture the continuum emission and a single Gaussian centered on the O VII line. The authors noted that the background subtraction process correctly removed the known instrumental SiK$\alpha$ line at 1.74 keV so that there is no reason to believe that the oxygen line is not a signature of the IPT or the moons. 

The authors suggested that much of the soft X-ray peak, particularly that observed at lower energies, may be the tail of the FUV emission studied by \textit{EUVE} \cite{gladstone_recent_1998}. To identify the source of the oxygen line, the authors calculated whether the solar flux would be sufficient to generate the observed signature and found that solar induced fluorescence would produce approximately one order of magnitude less photons than observed. The authors similarly calculated the charge exchange emission from oxygen but found that this produced emission 4 orders of magnitude below what was needed. However, recent work has determined that the required ion energies for charge exchange are significantly lower than initially thought \cite{houston_jovian_2020}, so that combined with recent particle measurements the available ion fluxes increase by 4-5 orders of magnitude \cite{garrett_jovian_2011,mauk_energetic_2004}. The observed emissions may therefore be feasible through charge exchange [E. Roussos, priv. comm.], but further modelling is required to test this calculation.

\subsection{X-rays from the Radiation Belts}
\label{sec:radbelts}
Earth, Jupiter, Saturn, Uranus and Neptune all possess robust radiation belts, suggesting that these may be a universal property of strongly magnetised planets. Radiation belts contain high intensities of trapped energetic particles. Of the planets, Jupiter possesses the most intense radiation belts with 2 orders of magnitude higher intensities of $\sim$10 MeV electrons than the other planets \cite{mauk_electron_2010}. For Jupiter, these belts begin within 1 $R_J$ of the cloud tops and continue to the orbit of Europa, with ions with energies of 100s of MeV, and potentially GeV, detected between 2-4 $R_J$ \cite{kollmann_jupiters_2021}.

\begin{figure}
    \centering
    \includegraphics[width=0.8\textwidth]{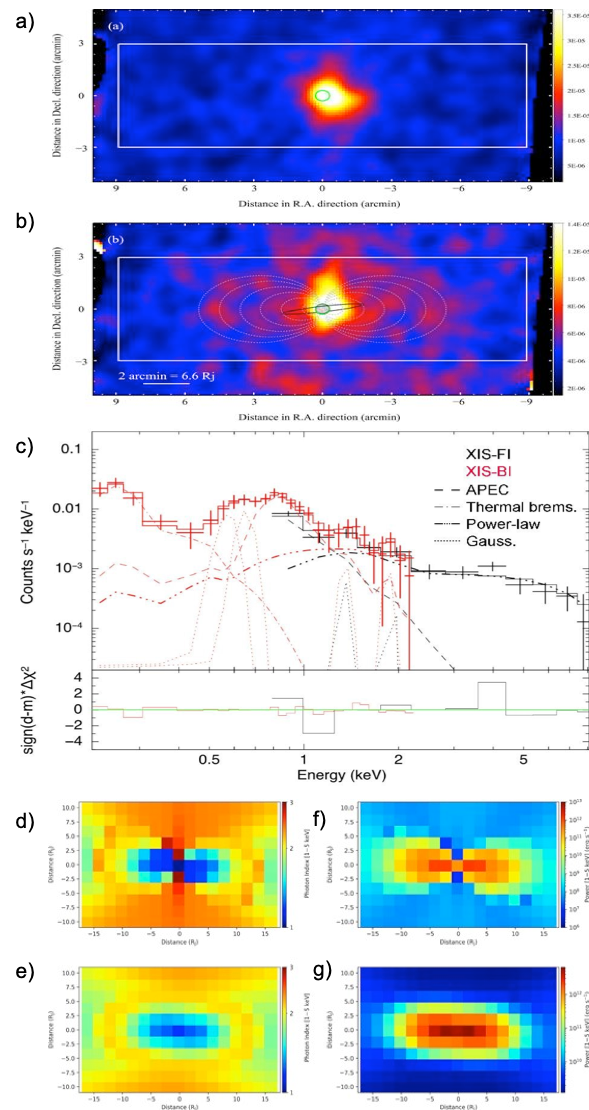}
    \caption{\textit{Suzaku XIS} Jupiter images in the (a) 0.2–1 keV (Back Illuminated - BI) and (b) 1–5 keV (Front Illuminated - FI) bands. The circle indicates the position and size of Jupiter (39''). Grey lines show magnetic field lines with equatorial crossings of 2, 4, 6, and 8 $R_J$. The black line is the path traced by Io. c) \textit{XIS} FI (black) and BI (red) background-subtracted spectrum from the extended emission region shown in b for an observation in 2012. The spectrum is overlaid with the best-fitting APEC + thermal bremsstrahlung + power-law + Gaussian line model. Residuals are shown in the lower panel. Modelled Inverse Compton emission from the Divine-Garett-model \cite{divine_charged_1983} of particle distributions in a meridian surface (d and f) and with three-dimensions (e and g) integrated in the direction of the line of sight for photon indexes of inverse-Compton 1-5 keV X-ray spectra (d and e), and powers of inverse-Compton 1-5 keV X-rays (f and g). Figures from \cite{ezoe_discovery_2010,numazawa_suzaku_2021}.}
    \label{fig:ezoeradbelts}
\end{figure}

Three \textit{Suzaku X-ray Imaging Spectrometer (XIS)}\cite{mitsuda_x-ray_2007}$\sim$160 ks observations of Jupiter taken in February 2006, January 2012 and April 2014 have each independently revealed a diffuse hard X-ray emission beyond the Jovian disk over distances of 12-21 R$_J$ by 3.5-8.2 R$_J$ (see example boxes in Figure \ref{fig:ezoeradbelts}). This diffuse emission is co-located with the radiation belts and Io Plasma Torus \cite{ezoe_discovery_2010, numazawa_suzaku_2019,numazawa_suzaku_2021}. These observations span solar minimum and solar maximum and it is found that while Jupiter's X-ray luminosity varies by a factor of three during this time (when scaled for the Jupiter-Earth distance), the diffuse emission only varies by a factor of $\sim$0.5, with a luminosity at Jupiter of 4.4-6.9 x 10$^{15}$ erg s$^{-1}$.

Figure \ref{fig:ezoeradbelts}a) and b) shows the observations of this emission from April 2014 \cite{numazawa_suzaku_2019}. Figure \ref{fig:ezoeradbelts}a) shows the soft X-ray component (0.2 - 1 keV), for which the spread of the emission is consistent with a point source combined with the instrument PSF - suggesting that this is almost entirely emission from Jupiter. In contrast, Figure \ref{fig:ezoeradbelts}b) shows the 1-5 keV emission, which is more consistent with a point source plus a uniform emission from a surrounding ellipse. 

Figure \ref{fig:ezoeradbelts}c) shows the spectrum from the region during 2012, with spectral lines between 0.2-1.3 keV well-matched to a combination of scattered solar emission and auroral charge exchange emission from Jupiter. From 1-5 keV,  the extended hard X-ray component is best represented by a flat continuum (photon index $\Gamma = 0.7-1.2$), suggestive of a non-thermal mechanism from either synchrotron emission, bremsstrahlung emission or inverse-compton scattering\cite{ezoe_discovery_2010}. At Io's orbit the magnetic field strength would require TeV electrons to generate synchrotron emission, so the authors ruled this possibility out. Later evidence further supported the exclusion of X-ray synchrotron with the brightness of the radio synchrotron seen to vary anti-phase with the solar cycle \cite{han_investigating_2018} - a pattern not observed in the X-ray emissions. The authors also rejected bremsstrahlung by comparing the cross sections for K-shell ionization of sulphur and noting that the resulting 2 keV spectral line should have been observable, but was not. Consequently, they concluded that inverse-Compton scattering between ultra-relativistic (several $\sim$10 MeV) electrons and visible solar photons was the most likely candidate. The characteristic energy of the inverse-Compton scattered photon is given by:
\begin{equation}
\sim3 keV (\frac{E_{ph}}{1.4 eV}) (\frac{E_e}{20 MeV})
\end{equation}
\\
where $E_{ph}$ and $E_e$ are the energies of the solar photons and the relativistic electrons before the scattering process respectively. Figure \ref{fig:ezoeradbelts}d-g) shows a series of models that combine a particle distribution model for Jupiter's magnetosphere \cite{divine_charged_1983} with calculations of the inverse-Compton scattering. While the photon index and spatial distribution from the model are reasonably similar to observations, the resulting fluxes are $\sim$100 times lower than those observed. The reason for this discrepancy is that the 10-50 MeV electron number density would need to be 7-50 times higher at the time of the \textit{Suzaku} observations than the 0.0007 and 0.0001 cm$^{-3}$ values measured by \textit{Pioneer} and \textit{Voyager} and used in the \textit{Suzaku} studies \cite{ezoe_discovery_2010,numazawa_suzaku_2021}. There is some evidence to suggest that the particle distribution model used to produce the inverse-Compton emission does underestimate the particle densities by a factor of 10 \cite{bolton_divine-garrett_2001}, but further work is needed to fully determine the source of the discrepancy.

Unfortunately, \textit{Suzaku} ceased operation in 2015, so no further observations can be acquired, but analysis of the megaseconds of Jupiter observations in the \textit{Chandra} and \textit{XMM-Newton} archives may provide valuable new studies of the radiations belts.

\section{\textit{X-ray Observations of the Galilean Satellites: Io, Europa, Ganymede and Callisto}}

Since their discovery in the early 1600s, Jupiter's Galilean satellites have been formative in our understanding of the Universe; sparking the creation of the Copernican view of the solar system. These icy worlds will be a key focus of space agency attention over the coming decades, with the ESA \textit{JUICE} and NASA \textit{Europa (Clipper)} missions due for launch in the coming year/s. Understanding the elemental and chemical abundances of these moons and how they interact with Jupiter's extreme radiation environment will be critical to understanding the formation and astrobiological potential of these worlds.

\begin{figure}
    \centering
    \includegraphics[width=\textwidth]{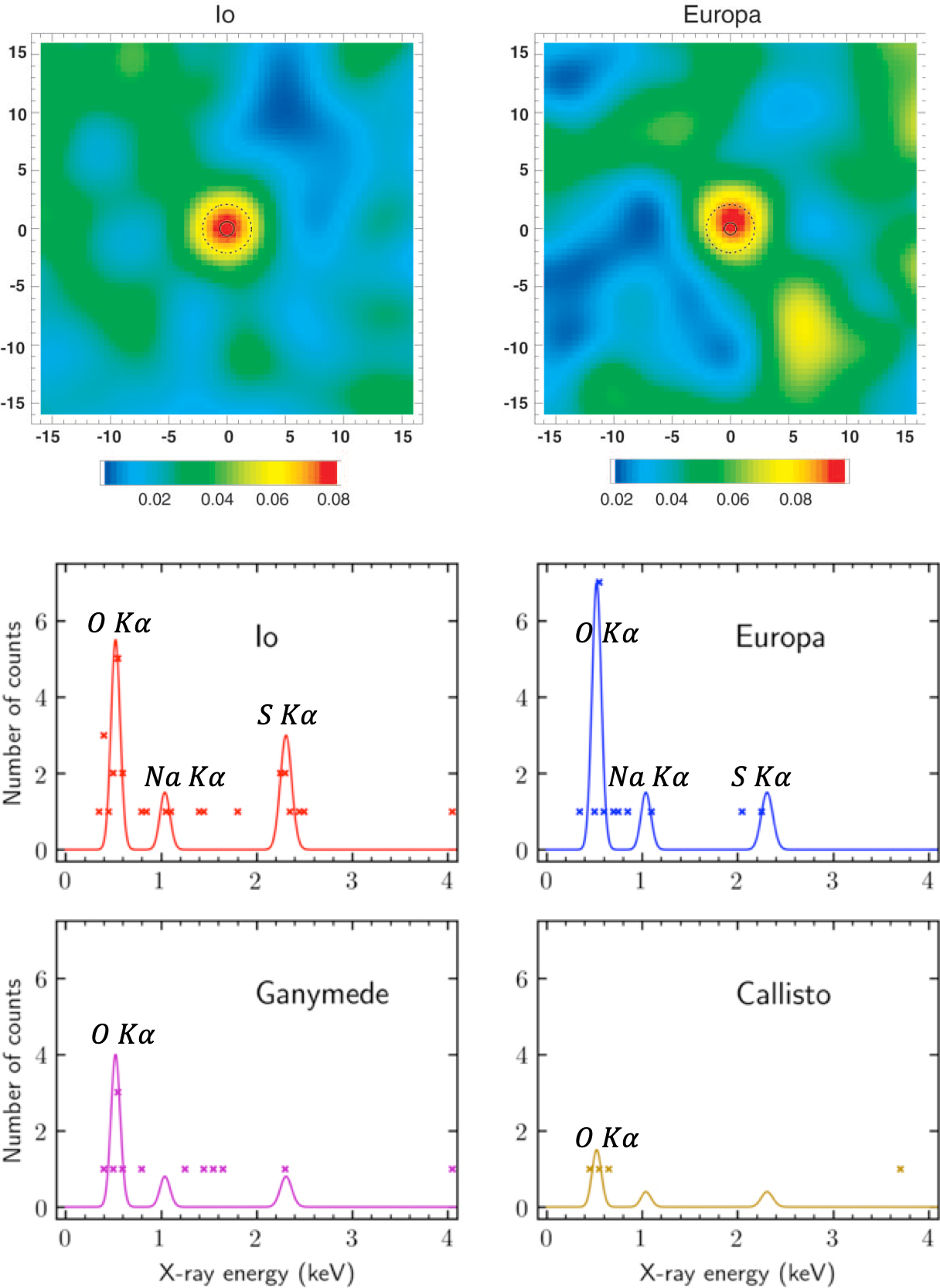}
    \caption{Top 2 panels: \textit{Chandra ACIS} images of Io and Europa ($250 eV <E <2000 eV$), smoothed by a 2D Gaussian ($\sigma = 2''.46$). The axes are in arcsec (100' = 2995 km), and the scale bar is in units of smoothed counts per image pixel. The solid circle shows the size of the satellite. Lower 4 panels: X-ray spectra of each of the Galilean moons, using 50 eV energy bins, overlaid with Gaussians at the positions of the O K-$\alpha$ (0.525 keV), Na K-$\alpha$ (1.041 keV), and S K-$\alpha$(2.308 keV) lines. Top 2 panels from \cite{elsner_discovery_2002}. Lower 4 panels from \cite{nulsen_x-ray_2020}.}
    \label{fig:elsnernulsen}
\end{figure}

The extended fields of view of \textit{Chandra} and \textit{XMM-Newton} mean that the orbits of the Galilean satellites are often within the field for Jupiter observations. The discovery of the X-ray emissions from Io and Europa occurred with the very first \textit{Chandra} observations of Jupiter in 1999 and 2000 \cite{elsner_discovery_2002} (detected telescope fluxes of $4 x 10^{-16}$ ergs s$^{-1}$ cm$^2$ (2 MW) for Io and $3 x 10^{-16}$ ergs s$^{-1}$ cm$^2$ (1.5 MW) for Europa), while Ganymede and Callisto were confirmed subsequently using analysis of several ACIS observations\cite{nulsen_x-ray_2020}.  Figure \ref{fig:elsnernulsen}a) shows \textit{Chandra} observations of the moons. 

Figure \ref{fig:elsnernulsen}b) shows how the photons are distributed by energy. Gaussians for the O, S and Na K-$\alpha$ lines are shown overlaid on the spectrum to illustrate the location of these features\cite{nulsen_x-ray_2020}. An X-ray fluorescence line occurs when an energetic process (often a photon or particle collision) ejects an electron from the innermost shell (the K-shell) of an atom. When this happens, another electron drops into the level to take the missing electron's place. The change in energy levels of the new K-shell electron produces a photon with a characteristic energy. The photon-energy acts as a finger-print signature for the element producing it, enabling remote identification of the elemental composition of an object. Depending on the energy level configuration, for some elements X-rays can also be produced from L-shell fluorescence transitions and fluorescence can also characterise isotopic and molecular composition.

Analysis of the X-ray spectrum shows that Io has oxygen K-$\alpha$ (0.525 keV) and sulphur K-$\alpha$ fluorescence lines (2.308 keV) and also a broad continuum of emission (Figure \ref{fig:elsnernulsen}). Europa is strongly peaked at the neutral oxygen fluorescence line. Ganymede has an oxygen fluorescence line and a broad continuum of emission. Unfortunately too few counts have been detected from Callisto to date for modelling, but the observed X-rays are observed to cluster at the oxygen fluorescence line energy. There are also hints of sodium K-$\alpha$ lines (1.041 keV) for the moons\cite{nulsen_x-ray_2020}. Unfortunately, the number of photons detected so far from each moon with \textit{ACIS} is too small to explore variation in composition between the leading and trailing hemispheres, which are known to be very different for e.g. Europa because of the differing levels of interaction with magnetospheric plasma.

While X-ray fluorescence at Mercury or the Moon is produced through solar photons and solar wind particle impacts, the fluxes for solar sources of energetic impacts are substantially lower at Jupiter's orbit ($\sim$ 5 AU). Instead, Io and Europa experience energetic particle fluxes from the Io Plasma Torus and radiation belts that are respectively $\sim$20 and 110 times higher than the solar photon fluxes at the moons\cite{elsner_discovery_2002}. Therefore, the two most likely processes to generate the observed spectra from the moons are thick-target bremsstrahlung emission \cite{pella_analytical_1985} from precipitating electrons and/or particle induced X-ray emission (PIXE \cite{johansson_particle-induced_1995}) from precipitating ions. Which process dominates may differ for each moon, depending on a variety of factors including the energetic particle environment for the moon and its atmospheric density \cite{paranicas_ion_2002}. 

PIXE is commonly used on Earth for a variety of industrial processes that require a non-destructive method for determining the elemental composition of an object, so that the physics is well characterised with existing models. One example model is the GUPIX code (http://pixe.physics.uoguelph.ca/gupix/main/), which offers a software package that can be applied to explore the PIXE emission from the Galilean satellites \cite{blaauw_2000_2002, nulsen_x-ray_2020}. For the Galilean moons the PIXE emissions can be defined as:

\begin{equation}
    F_L = \pi(R_{moon}/d_{Earth})^2 \times \pi N_0 \int_{E_l}^{E_u} (E_{ion})^{-\alpha} Y(S,E_L,E_{ion})dE_{ion}
\end{equation}

where, $F_L$ is the X-ray flux, of a characteristic emission line $L$ with energy $E_L$ in units of photons $cm^{-2} s^{-1} ppt^{-1}$ (ppt = part per thousand). $R_{moon}$ and $d_{Earth}$ are the radius of the chosen moon and distance between that moon and the observer, respectively. $E_{ion}$ is the proton energy in keV. $N_0$ is the normalization and $\alpha$ is the spectral energy index of the proton spectrum (\cite{nulsen_x-ray_2020} take this to be a power law with $N_0$ in units of protons $cm^{-2}s^{-1} sr^{-1} keV^{-1}$), $E_l$ and $E_u$ are the lower and upper proton energies (\cite{nulsen_x-ray_2020} take this to be 0.1 and 10 MeV), $E_L$ is the characteristic line energy of the relevant element, $S$ is the surface composition with the fraction of each element in the surface material given in units of ppt by mass, and Y is the PIXE yield in units of photons sr$^{-1}$ ppt$^{-1}$ ion$^{-1}$.

\begin{figure}
    \centering
    \includegraphics[width=\textwidth]{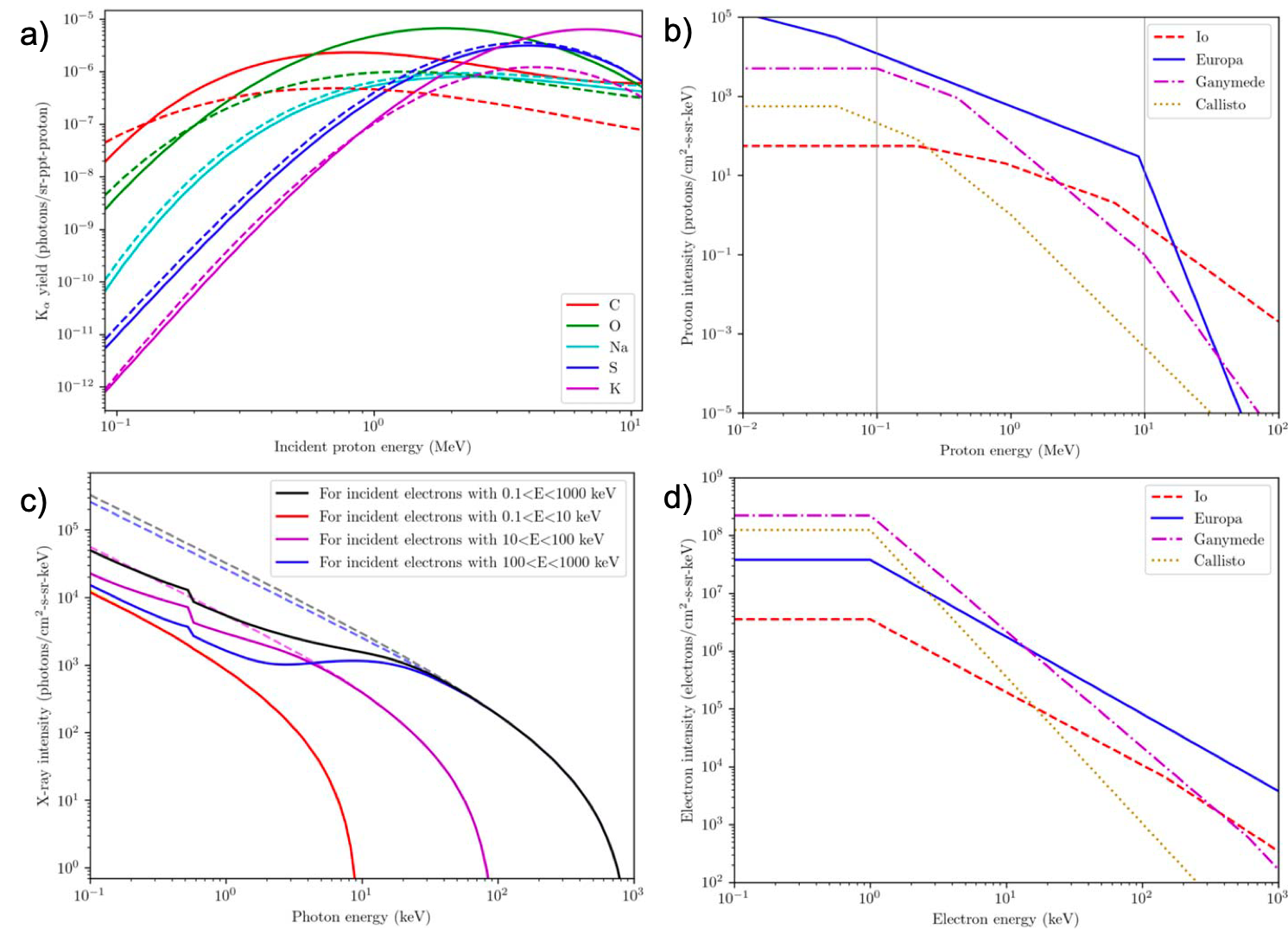}
    \caption{a) Io (dashed lines) and Europa (solid lines) yields of Particle Induced X-ray Emission (PIXE) (photons/sr-ppt-proton) for various proton energies, with the respective compositions assumed to be ices of SO$_2$ and H$_2$O \cite{nulsen_x-ray_2020}. b) For the orbital locations of each Galilean moon, piecewise continuous fits to measured proton intensities. The vertical grey lines at 0.1 and 10 MeV indicate the bounds of the proton energies used to calculate the X-ray fluxes. c) Europa's thick-target bremsstrahlung\cite{pella_analytical_1985} predicted X-ray intensities for various incident electron energy ranges. Calculations include self-absorption, with the dashed lines highlighting what the flux would be if self-absorption was not accounted for.  d) Same as b, but for electron intensities. Figures from \cite{nulsen_x-ray_2020}.}
    \label{fig:nulsen2}
\end{figure}

Taking detailed compositions and precipitating particle populations from the literature, it is then possible to generate expected PIXE fluxes and to compare these with the observed data. Figure \ref{fig:nulsen2}a shows the resulting K-$\alpha$ line yields for Io and Europa for a range of proton energies and Figure \ref{fig:nulsen2}b shows proton intensities at the moons. Variation in emission from the moons can depend on the composition of the surface observed and the energetic particle fluxes incident at the moon.  

In contrast with the PIXE models for ion bombardment, which only generate the line emissions, the electron bombardment will produce continuum emissions and line emissions. For elements with atomic numbers lower that 13, the line emissions will contribute 10-20\% of the emission, but for elements with atomic numbers above 13 (such as sulphur) the line emissions can be up to 50\% of the observed emissions. However, electrons may inject their energy too deep into the surface for the fluorescence photons to escape \cite{elsner_discovery_2002,elsner_x-ray_2005}. Consequently, an electron bombardment dominated X-ray emission would be continuum emission-dominated for the water ice moons Europa and Ganymede, but may have contributions of continuum and line emission for the sulphur rich surface of Io. The X-ray intensity from bremsstrahlung is given by:

\begin{equation}
    I(E) = K Z (E_0/E - 1)
\end{equation}

where $E_0$ and $E$ are the incident electron and radiated photon energies, respectively, in keV, K is a constant (2.2 x $10^{-7}$ photons keV$^{-1}(e^-)^{-1} $sr$^{-1}$), and Z is the atomic number of the target material \cite{markowicz_composition_1984}. However, since the moons are not made of a single element with only one atomic number, it is common practice to use an effective atomic number \cite{markowicz_composition_1984} that aggregates the atomic weights of the target material. The resulting bremsstrahlung X-ray intensities for Europa are shown in Fig \ref{fig:nulsen2}c) with associated electron intensities in Figure  \ref{fig:nulsen2}d).

The emission from Io is well-matched to expected spectra from electron bremsstrahlung models. For Europa, the observed emission exceeded the observations for PIXE or bremsstrahlung, but the count-rates were very low so it remains unknown which process dominates. For Ganymede the observed continuum emission favours a bremsstrahlung model, while Callisto's X-ray signal is too low for modelling\cite{nulsen_x-ray_2020}. 

The majority of these studies utilised the \textit{Chandra ACIS} instrument, whose  unique combination of high spatial resolution and spectral resolution is invaluable for the moons. Unfortunately, with the contaminant build-up on \textit{Chandra ACIS}\cite{plucinsky_complicated_2018} at the soft energy range most relevant for the moons, it has not been possible to observe them with ACIS since 2011. While it is possible to detect the moons with HRC-I \cite{elsner_discovery_2002, nulsen_x-ray_2020}, the signal is more dominated by the background emission and offers no energy resolution with which to inspect the spectrum. For example, for Io, from 585 ks of Chandra HRC-I observations 123 photons were detected from the moon, but 105 of these are expected to be background emission \cite{nulsen_x-ray_2020}. This means that it is challenging to use the HRC data to, for instance, track variation with orbital phase.

Arguably, the greatest value that X-ray observations can contribute to our understanding of Jupiter's moons is the characterisation of their elemental composition through the fluorescence lines. No other waveband is capable of providing these elemental constraints, particularly on trace species such as chlorine and potassium, which are key to understanding Europa's sub-surface ocean. To probe these constraints is likely to require the next generation of X-ray instrument.

\section{\textit{Future Observations}}

\begin{figure}
    \centering
    \includegraphics[width=\textwidth]{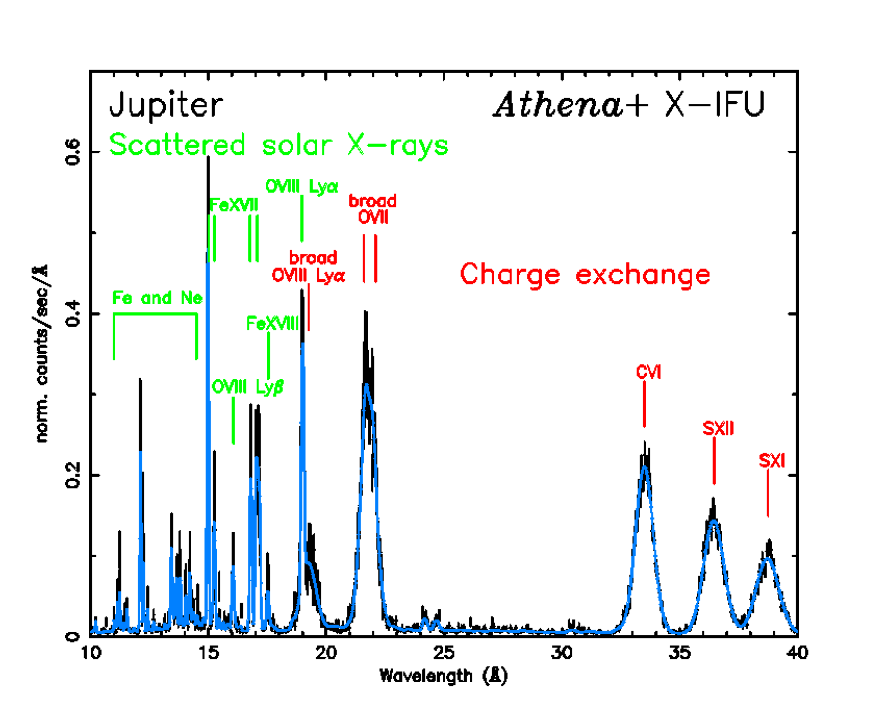}
    \caption{Simulated 20 ks exposure ATHENA X-IFU spectrum for the model shown in Fig \ref{fig:gbr2004xmm}. The only relevant background contributions would be from particles, which would be negligible and is therefore not included. Figure from \cite{branduardi-raymont_hot_2013}.}
    \label{fig:gbrATHENA}
\end{figure}

With the launch of ESA's \textit{ATHENA} (Advanced Telescope for High ENergy Astrophysics) X-ray Observatory in the early 2030s, we will enter a new era of X-ray astronomy \cite{nandra_hot_2013}. The X-IFU (X-ray Integral Field Unit) \cite{barret_hot_2013} will enable collecting high-signal high-energy-resolution spectra from Jupiter. For example, figure \ref{fig:gbrATHENA} shows a modelled 20 ks X-IFU observation of Jupiter. On an observation-by-observation basis this will enable the characterisation of the particle precipitation at Jupiter and the relative contributions of Iogenic vs solar wind ions. In comparison with approach phase and magnetospheric measurements by ESA's \textit{JUICE} and NASA's \textit{Europa (Clipper)} missions, this will enable revolutionary studies of the global dynamics, addressing key questions of to what extent Jupiter is open to the solar wind and what conditions enable this for rapidly rotating giant magnetospheres. NASA's proposed next generation X-ray observatory, \textit{Lynx} \cite{gaskin_lynx_2019}, would enable high spatial resolution to accompany the high count-rates for the aurorae, pin-pointing precise structures and morphologies that relate to changing conditions \cite{snios_x-rays_2019}.

Similarly, at high energies such instruments will map variation across Jupiter's radiation belts and Io Plasma Torus and study how this changes with time. Again, coupling this with in-situ observations will connect the chain of causality that produces global, remotely-detected changes in these large-scale plasma structures. For the Galilean satellites, the order of magnitude increase in effective area will permit firm detections of fluorescence emissions and consequently characterise the elemental composition of the surfaces of the moons. Enhanced count-rates may permit comparisons between leading and trailing hemispheres (observed at different intervals in their orbit), distinguishing how precipitating particles drive the emissions and what chemistry is available for potential astrobiology across the moons.

While these observatories will be transformative for studies of all of the X-ray emitting structures in Jupiter's magnetosphere, an orbiter or lander is needed to truly map the elemental composition of the moons. Due to proximity alone, the flux would be
6–7 orders of magnitude larger \cite{nulsen_x-ray_2020}, such that the presence of trace elements in small concentrations could easily be measured in short exposure times. Depending on proximity to the moon, such abundances could be mapped to up to tens of meter resolution. Fortunately, recent advances in X-ray instrumentation enable the possibility of such instrumentation.

\subsection{Forthcoming and Proposed In-situ X-ray Instruments}

Since the Apollo era, there has been a small heritage of X-ray instrumentation on in-situ missions to the inner solar system, led by missions such as \textit{MESSENGER} and \textit{Chandrayaan}. In the coming years, this science legacy will proliferate: the \textit{MIXS} \cite{fraser_mercury_2010} instrument on \textit{ESA-JAXA's BepiColombo} is set to study the composition and formation of Mercury through X-ray fluorescence, and the Soft X-ray Imager on the \textit{ESA-CAS SMILE} \cite{SmileGBR} mission will obtain the first global X-ray images of the magnetosheath of Earth - providing invaluable global insights into the relation between the terrestrial magnetosphere and space weather. 

A soft X-ray instrument has never visited the outer solar system, despite the available wealth of profound new scientific findings on aurora, magnetospheric interactions and elemental composition and formation of moons, rings and atmospheres. To operate in the Jupiter environment in-situ X-ray instruments face two key challenges: 1. the harsh radiation environment of Jupiter provides both noise and potentially damaging particle impacts for an instrument, 2. Jupiter is a bright optical source and so requires a method to prevent contamination from infrared, visible and UV photons. Recent advances in X-ray instrumentation such as Micro-Pore Optics, miniature X-ray optics and radiation tolerant detectors open the possibility of compact, lightweight, X-ray instrumentation perfectly suited for Jupiter science. Several concepts have been proposed and are currently being investigated for Jupiter missions. Here, we briefly highlight a few potential future instruments for Jupiter to show the feasibility and pioneering new directions for this dynamic and rapidly evolving field.

\begin{figure}
    \centering
    \includegraphics[width=\textwidth]{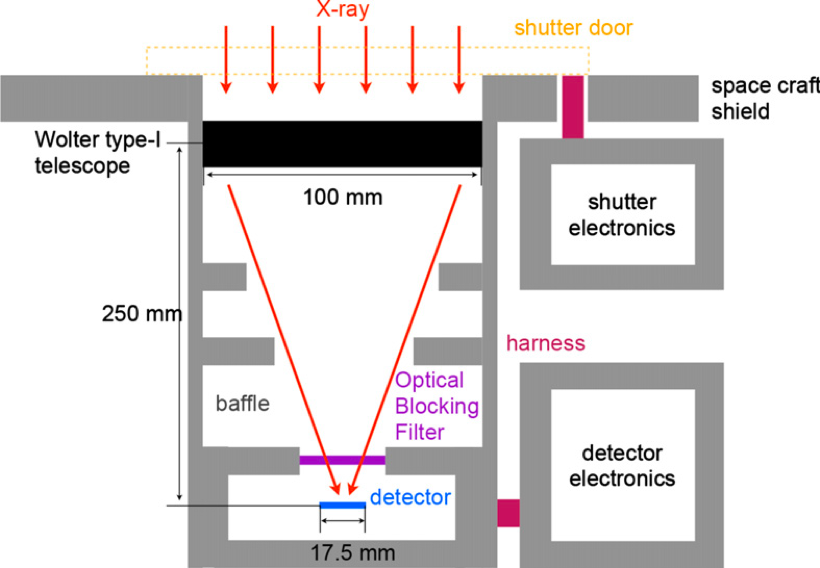}
    \caption{The design concept for the JUXTA in-situ Jupiter X-ray Instrument on the JAXA JMO mission. Figure from \cite{ezoe_juxta_2013}.}
    \label{fig:JUXTA}
\end{figure}

\textit{JUXTA}, the Jupiter X-ray Telescope Array \cite{ezoe_juxta_2013}, was a concept instrument for the JAXA \textit{Jupiter Magnetospheric Orbiter (JMO)}\cite{tao_magnetospheric_2011, sasaki_japanese_2011}. Unfortunately the mission was not implemented. \textit{JUXTA} consisted of a lightweight X-ray telescope and a radiation-hard semiconductor pixel detector. The instrument would offer 0.3-2.0 keV coverage (at $<$100 eV resolution) ideal for the charge exchange emissions from the aurorae, and exosphere and for capturing fluorescence lines from the moons' surfaces (while this does not capture Sulphur K lines, it does capture L lines). The combination of the instrument's 5' spatial resolution and 3 cm$^2$ effective area at 0.6 keV with the proposed mission orbital radius (30 $R_J$) would have enabled high spatial resolution and sensitivity - the equivalent resolution and sensitivity for an Earth-orbiting telescope would be 1'' and 24 m$^2$, respectively, obtaining $\sim$200-1000 counts per second from Jupiter's aurorae. The proposed telescope concept is shown in Figure \ref{fig:JUXTA}. The telescope can fit within a 10 kg payload cap through the virtue of ultra lightweight micropore optics \cite{ezoe_mems-based_2014, ogawa_iridium-coated_2013}. For the detector, a Jupiter in-situ X-ray instrument likely needs to avoid using traditional CCD's, which would be degraded by the intense fluxes of energetic particles in the Jovian magnetosphere \cite{elsner_x-ray_2005}. Ideally, a replacement detector would not require charge clocking or large distances for charge packets to travel, since otherwise the high fluxes mean that these are likely to result in additional charge build-up during read-out. \textit{JUXTA} favoured a CMOS (Complementary metal-oxide-semiconductor) detector, which offers a more radiation hardy alternative \cite{kenter_development_2012} and included electronics that enabled fast readout \cite{ezoe_mems-based_2014, ogawa_iridium-coated_2013}. While CMOS detectors can be more tolerant of visible photon contamination, the brightness of Jupiter necessitates an optical blocking filter, which reduces optical contamination to a negligible level.  Electron fluxes are the dominant source of instrument damage \cite{elsner_x-ray_2005,ezoe_juxta_2013}. While the dose rate is not damaging for the instrument, in Jupiter's plasma disk, the count-rates from the particle and $\gamma$-ray background are comparable to the planetary signal, even with shielding, making it difficult to distinguish signal from background. However, in the high latitude regions - away from Jupiter's plasma disk - count-rates due to energetic particle events are orders of magnitude below the signal from Jupiter, and would not be problematic. Consequently, while JUXTA could not have collected useful observations from within the Jovian plasma disk, it would have been safe from these radiation doses, and could have delivered high value science whenever the spacecraft was outside of the plasma sheet. This well-conceived instrument was ideally equipped to address the diverse range of science available in the Jovian system.

While \textit{JUXTA} offered a general purpose Jupiter X-ray instrument, the NASA-funded study X-ray Mapper of Icy Moon Elements (X-MIME), focused on a dedicated instrument for the Galilean satellites \cite{elsner_x-ray_2005}. This study identified a feasible orbiting telescope at one of the moons, with 0.2-8 keV coverage and 150 eV resolution which would provide detailed maps of the elemental composition of the surface of one moon, while being able to image the other moons and Jupiter. It highlighted Ganymede as the preferable location due to the more benign radiation environment. During approach, a 6$^\circ$ field of view at distances of 200 $R_J$ or greater would enable continuous observation of the X-ray emission from the IPT and possibly also any neutral clouds of material in the system, with the capacity to identify elemental composition through fluorescence.  On approach, a 6$^\circ$ field of view would also enable Io to be observed in concert with Jupiter continuously until the spacecraft was 120 $R_J$ from the planet. Preceding the recent developments of CMOS detectors, two alternatives would be: pixellated silicon PIN diode arrays or pixellated silicon drift detectors \cite{elsner_x-ray_2005,struder_european_2001}. 

The heritage of these studies has continued with concept studies underway for X-ray fluorescence instrumentation to study the chemical composition of Europa \cite{kraft_x-ray_2018, tremblay_x-ray_2018}. In-situ instruments currently being studied could measure the presence of Na, Mg, or Cl in the surface
of Europa to a level of 1 ppt by mass in about 15 minutes. Depending on orbital proximity to the moon, it may be possible to map these to regions of 10s of meters \cite{nulsen_x-ray_2020}. Furthermore, were these to be used on a nominal \textit{Europa Lander} mission, then they could map elemental surface composition on scales of less than meters \cite{kraft_x-ray_2018, tremblay_x-ray_2018}. By probing variation in the particle fluxes at the moons, through these fluorescence emissions, it may also be possible to track induced currents from the sub-surface oceans [N. Achilleos, priv. comm] \cite{saur_search_2015}. 

In 2021, NASA's heliophysics division funded the concept study of the \textit{COMPASS} spacecraft (formerly \textit{JUGGERNOT}). This spacecraft would seek to study in-situ Jupiter's radiation belts by using a suite of plasma and magnetic field instruments and an X-ray instrument [Roussos et al., in review; Clark et al., in review]. The 6 kg X-ray instrument's prime mission objective would be to provide global long-timescale analysis of the radiation belts. The instrument would be based on the terrestrial radiation belt AEPEX design, which will soon fly on a terrestrial 6U CubeSat,  offering 50-300 keV energy range and 10s resolution \cite{marshall_aepex_2020}.  Work on this concept is underway at the time of writing this chapter and the instrument design may change substantially.

\newpage

\section{\textit{Summary}}

Four decades of remote X-ray observations have offered fundamental insights into Jupiter's high energy space environment. Table \ref{table:summary} shows a non-exhaustive summary of the main sources of X-ray emission from the Jovian system. Most recently, the coupling of \textit{Juno} in-situ measurements with \textit{Chandra, XMM-Newton} and \textit{NuSTAR} observations has begun to reveal the physical processes that generate Jupiter's auroral emissions. For the coming decade, this rich heritage dataset will continue to revolutionise our understanding of the system. \textit{ATHENA} and, nominally, \textit{Lynx} coupled with in-situ measurements by \textit{JUICE} or the \textit{Europa (Clipper)} mission, will level-up this understanding and undoubtedly reveal new, potentially yet to be imagined, emissions from the system. The future is very bright for this field.

\begin{table}
  \caption{Observed (upper table) and Expected (lower table) X-ray Sources in the Jovian System}
  \centering 
    \begin{tabular}{cccc}
Source & Observed Energy &  Dominant Mechanism & \textbf{Confirmed} \& Expected Species\\
    \hline\hline\\
    Jupiter's Polar Aurorae & $<0.9^A$ keV & Charge Exchange (CX) & \textbf{O}, S, C, N, Mg\\
    \\
    Jupiter's `Main'$^B$ Aurora & 2-20 keV & Bremsstrahlung$^C$ & \textbf{Electrons (e$^-$)}\\
    \\
    Upper Atmosphere & $>0.2$ keV & Scattered Solar Photons & \textbf{Fe, Ne, Mg, Si, (e$^-$)}, O, N, C \\
    \\
    Radiation Belt Emissions & 1-5 keV & Inverse Compton Scattering & \textbf{$>$10 MeV e$^-$} \\
     \\
    Io Plasma Torus & 0.2-1 keV & CX, Bremsstrahlung & \textbf{O, e$^-$}\\
    \\   
    Galilean Moons & 0.2-3 keV & Fluorescence$^D$, Bremsstrahlung & \textbf{O, S, e$^-$}, Na, K, Ca\\
    \\ \hline\hline\\
    Radiation Belt Precipitation & 0.2-50 keV & CX, Bremsstrahlung & O, S, e$^-$\\
    \\
    Exosphere Emissions & 0.2-1.5 keV & Iogenic CX, Solar Wind CX & O, S, C, N, Mg \\
    \\ \hline\hline\\
    
\end{tabular}
$^{A}$ Some Mg and S CX lines exist above 1 keV, but these are yet to be confirmed in spectra.\\ 
$^{B}$ Definition stems from UV, but the region is dimmer than the ion aurorae for the X-rays.\\
$^{C}$ Thermal and non-thermal components.\\
$^{D}$ From Particle Induced X-ray Emission (PIXE) and/or Thick-Target Bremsstrahlung.
\label{table:summary}
\end{table}

However, there are a variety of structures for which X-ray observations are the key to addressing fundamental questions, but for which the signal and/or spatial resolution is too low from Earth-orbit observations. Uncovering the global behaviours, sources and sinks of the radiation belts relativistic populations or mapping the elemental composition of the surfaces of the Galilean moons will require an in-situ X-ray instrument. Such an instrument would also provide a step-change in our understanding of the other sources in Table \ref{table:summary}. Recent advances in technology enable this possibility and such an instrument would offer unique capabilities for any heliospheric or planetary mission to the Jovian magnetosphere; providing a vast discovery space, from questions of the prevalence of life to the fundamental acceleration processes in the cosmos. 

The science cases are rich, diverse and compelling. The technology is ready. The stage is set for a soft X-ray imager for the outer planets; it simply remains a question of which agency or organisation will seize the scientific opportunities presented.

\end{document}